\documentclass[final,5p,times,twocolumn]{elsarticle}

\usepackage{graphicx} 
\usepackage[inline]{enumitem}
\usepackage{mdframed}
\usepackage{comment}
\usepackage{booktabs}
\usepackage{float}
\usepackage{multirow}
\setlist[enumerate,1]{label=\textit{\alph*)}}
\usepackage{amsmath,amsfonts}
\usepackage{textcomp}
\usepackage{flushend}
\usepackage{array}
\usepackage{multirow}
\usepackage[inline]{enumitem}
\usepackage{booktabs}
\usepackage{float} 
\usepackage{tcolorbox}
\usepackage{multirow}
\usepackage{array}
\usepackage{url}
\usepackage{subcaption} 
\usepackage{caption}
\usepackage[hidelinks]{hyperref}
\usepackage{algorithm}
\usepackage{algpseudocode}
\usepackage{wrapfig, booktabs}
\usepackage{graphicx,xcolor,lipsum,multicol,caption}
\usepackage{subcaption}
\def\BibTeX{{\rm B\kern-.05em{\sc i\kern-.025em b}\kern-.08em
    T\kern-.1667em\lower.7ex\hbox{E}\kern-.125emX}}

\definecolor{codegreen}{rgb}{0,0.6,0}
\definecolor{codegray}{rgb}{0.5,0.5,0.5}
\definecolor{codepurple}{rgb}{0.58,0,0.82}
\definecolor{backcolour}{rgb}{0.95,0.95,0.92}

\usepackage{listings}
\usepackage{color}

\definecolor{dkgreen}{rgb}{0,0.6,0}
\definecolor{gray}{rgb}{0.5,0.5,0.5}
\definecolor{mauve}{rgb}{0.58,0,0.82}

\NewTColorBox{custombox}{o}{
    colback=gray!10,
	colframe=gray!10,
	left=1.0mm,
	right=1.0mm,
	top=1.0mm,
	bottom=1.0mm,
	fonttitle=\bfseries,
	arc=0mm,
	leftrule=0mm,
	rightrule=0mm,
	toprule=0mm,
	bottomrule=0mm,
	notitle,
	before=\par\medskip\noindent,
    IfValueTF={#1}{before upper={\textbf{#1: } }}{},
    parbox=false,
}
\definecolor{codegreen}{rgb}{0,0.6,0}
\definecolor{codegray}{rgb}{0.5,0.5,0.5}
\definecolor{codepurple}{rgb}{0.58,0,0.82}
\definecolor{backcolour}{rgb}{0.95,0.95,0.92}

\usepackage{listings}
\usepackage{color}
\usepackage{fancyvrb}
\definecolor{dkgreen}{rgb}{0,0.6,0}
\definecolor{gray}{rgb}{0.5,0.5,0.5}
\definecolor{mauve}{rgb}{0.58,0,0.82}

\newcommand{\revise}[2]{%
    \color{black}{#2}
}

\newcommand{\rqone}{How effective is the proposed approach for the Text-to-Testcase generation task?}
\newcommand{\rqtwo}{What is the effect of the fine-tuning and prompting step for the LLM-based text-to-testcase generation task?}
\newcommand{\rqthree}{What types of errors occur in the incorrectly generated test cases?}

\newcommand{\sectopic}[1]{\vspace{0.2em}\par\noindent{\textit{\bfseries #1}}}

\newcommand{\smallsection}[1]{{\bf #1}.\hspace{1mm}}
\newcommand{\ea}{\textit{et al.}}

\AtBeginDocument{%
  \providecommand\BibTeX{{%
    \normalfont B\kern-0.5em{\scshape i\kern-0.25em b}\kern-0.8em\TeX}}}

\begin{document}
\title{Enhancing Large Language Models for Text-to-Testcase Generation}
\author{Saranya Alagarsamy}
\author{Chakkrit Tantithamthavorn}
\author{Wannita Takerngsaksiri}
\author{Chetan Arora}
\author{Aldeida Aleti}

\begin{abstract}

\textbf{Context:} 
Test-driven development (TDD) is a widely employed software development practice that involves developing test cases based on requirements prior to writing the code. 
Traditional TDD is time-consuming, since writing test cases based on a given requirement is a complex task.
Although various methods for automated test case generation have been proposed, they are not specifically tailored for TDD, where requirements instead of code serve as input.

\textbf{Objective:} 
In this paper, we introduce a text-to-testcase generation approach based on a large language model (GPT-3.5) that is fine-tuned on our curated dataset with an effective prompt design. We aim to investigate the performance of LLMs-based text-to-testcase generation task based on two
contexts, i.e., when LLMs are leveraged by fine-tuning and prompting. 

\textbf{Method:} 
Our approach involves enhancing the capabilities of basic GPT-3.5 for text-to-testcase generation task that is fine-tuned on our curated dataset with an effective prompting design.
We evaluated the effectiveness of our approach using a span of five large-scale open-source software projects.  In total, we investigate 32 variations of eight LLMs (i.e., basic GPT-3.5-turbo,
Incoder, Starcoder, CodeT5, Bloom, CodeGemma, CodeLlama, and Gemini).

\textbf{Results:} 
Our approach generated 7k test cases for open source projects, achieving 78.5\% syntactic correctness, 67.09\% requirement alignment,  61.7\% code coverage, and 18.9\% mutation score which substantially outperforms all other LLMs.
Furthermore, our ablation study demonstrates the substantial performance improvement of the fine-tuning and prompting components of the GPT-3.5 model.

\textbf{Conclusions:} 
These findings lead us to conclude that fine-tuning and prompting should be considered in the future
when building a language model for the text-to-testcase generation task.

\begin{keyword}

TDD \sep Text-to-Testcase generation.
\end{keyword}



\end{abstract}
\maketitle
\newpage
\section{Introduction}

\revise{R1.1, R2.1}{Test-Driven Development (TDD) is a commonly used software development practice~\cite{beck2003test,madeyski2010test,barraood2021comparison} for designing and developing software, widely adopted by agile software development teams that prioritize creating test cases before the actual coding process.}
This methodology is structured around iterative cycles that involve the development of test cases based on a given requirement, the implementation of code to pass these test cases, and then refactoring to continuously improve the code.
TDD is claimed to significantly improve code quality and handle requirements better than 'more traditional, heavy weight predictive methods'~\citep{wasmus2007evaluation}.
The implementation of TDD led to a reduction in defects by approximately 50\%, in comparison to a similar system developed with ad-hoc unit testing methods~\citep{maximilien2003assessing}.
\revise{R1.1}{TDD empowers programmers to enhance code quality both before and after writing code. By regularly assessing testability and refactoring it to improve without executing the program, developers can ensure that production code is always test-ready.}
Despite these benefits, one of the primary drawbacks of this methodology is the time and effort required to write test cases. 
The initial investment in creating comprehensive and effective test cases is substantial and represents the largest overhead cost in the TDD process~\citep{beck2022test}. However, this investment is a critical component of TDD.

Various automated test case generation approaches have been proposed to help developers write test cases. 
Such approaches leverage various techniques, like random~\cite{pacheco2007randoop}, evolutionary algorithms~\citep{fraser2011evosuite}, and deep learning~\citep{tufano2020unit,alagarsamy2023a3test}.
However, existing test case generation approaches often require source code as input in order to generate test cases (Code$\rightarrow$Test), which makes them unsuitable for TDD.

The Large Language Models (LLMs), being advanced language models, are capable of performing various Natural Language Processing (NLP) tasks such as text generation, question answering, and classification. 
Specifically in the field of software engineering (SE), LLMs are increasingly being applied for tasks such as requirements engineering, code generation, vulnerability identification, test generation, and software maintenance. 
\revise{R2.2}{Large Language Models (LLMs) have significantly
excelled in automatic code generation \cite{chen2021evaluating, roziere2023code}, code completion~\cite{svyatkovskiy2019pythia} , automatic patch generation ~\cite{tufano2019empirical} ~\cite{chen2019sequencer}, comment generation ~\cite{hu2020deep}, and many others ~\cite{watson2020learning}. 
While LLMs demonstrate proficiency in various automated programming tasks, their application in automatic test case generation, particularly under Test-Driven Development (TDD) conditions, remains limited. To the best of our knowledge, there is a lack of in-depth investigation into the performance of LLMs for Text-to-Testcase generation tasks.

While other automatic test case generation approaches are proposed, none of these approaches are designed for TDD (i.e., taking requirement as input), are not able to generate test cases without code (i.e., mostly they take code as input).
Therefore, these limitations highlight the need
for an automated test case generation approach that takes a requirement as input and can generate
alternative test cases beyond the ones that exactly match the ground truth test cases.

As a result, it remains an open question whether LLMs can be effectively utilized for test case generation from textual descriptions. To the best of our knowledge, there is a lack of in-depth investigation into the performance of LLMs for Text-to-Testcase generation tasks. In this work, we aim to fill this gap by exploring how LLMs perform on Text-to-Testcase generation tasks. }

In this paper, we introduce a Text-to-Testcase generation approach, which involves a fine tuning component for the basic GPT-3.5 using our dataset specifically curated from open-source Java projects and an effective prompting design.
The model fine-tuning is aimed to  teach the LLM the mapping between the method description (\emph{text}), and the corresponding test (\emph{testcase}).
Once the basic GPT-3.5 model is fine-tuned, during the model inference phase, an unseen method description (i.e., the method description is referred to as the requirement) with the designed prompt is used as input into the fine-tuned GPT-3.5 to generate the corresponding test case without requiring the code.
Crucially, our approach differs from traditional methods by generating test cases directly from method descriptions without requiring the implemented code (method). 
This allows for the generation of test cases in the absence of actual code, a significant shift from existing practices.

In this work, we aim to investigate the performance of LLMs-based
text-to-testcase generation task based on two contexts, i.e., when LLMs are
leveraged by fine-tuning and prompting. 
In particular, we evaluate eight
LLMs (i.e., GPT-3.5~\citep{openai2023gpt}, Bloom~\citep{luccioni2022estimating}, CodeT5~\citep{wang2021codet5},  Incoder~\citep{fried2022incoder}, Starcoder~\citep{li2023starcoder}, CodeGemma~\citep{team2024codegemma}, CodeLlama~\citep{roziere2023code}, and Gemini~\citep{gemini2024}) fine-tuned on our newly curated benchmark dataset, consisting of 163k pairs of method descriptions and test cases.
We developed an evaluation dataset with 7K test methods
that span five large-scale open source software projects Apache Commons Lang~\citep{Apache}, JFreeChart~\citep{JFreeChart}, Apache Common CLI~\citep{ApacheCommonsCLI}, Apache Common CSV~\citep{ApacheCommonsCSv}, Google Gson~\citep{googleGson}.
For each model, we evaluate the quality of the generated test cases in four aspects: syntax correctness, requirement alignments,  code coverage and mutation score.
In summary, our objective is to address the following research questions (RQs):

\sectopic{RQ1. \rqone}\\
\smallsection{Results}
The proposed approach generates test cases that are 78.5\% syntactically correct and 67.09\% aligned with the requirements while achieving a code coverage of 61.7\%, and 18.9\% mutation score which substantially outperforms the baselines (basic GPT-3.5-turbo, Incoder, Starcoder, CodeT5, Bloom, CodeGemma, CodeLlama, and Gemini).


\sectopic{RQ2. \rqtwo}\\
\smallsection{Results}
The proposed fine-tuning component improves syntax correctness by 223\%, requirements alignment by 164\%, the coverage of the code by 153\%, and
mutation score improved by 273\%.
The proposed prompt design improves the syntax correctness by 124\%, the requirements alignment by 82\%, code coverage by 58\%, and mutation
score improved by 150\%.


\sectopic{RQ3. \rqthree}\\
\smallsection{Results}
We observed that 78.5\% of the generated test cases are syntactically correct, while 21.5\% of the generated test cases are incorrect.
Out of the incorrectly generated test cases (21.5\%) we observed that 11.3\% were associated with assertion errors, 2.4\% with value errors, 6.9\% with syntax errors, and the remaining 0.9\% with other types of errors.

Based on our results, we draw the following recommendation:

\textbf{The basic usage of ChatGPT may not produce desirable performance for the text-to-testcase generation task.}
Both fine-tuning and effective prompting design should always be considered when using a large language model. 
Our RQ1 shows that our proposed approach performs the best for the text-to-testcase generation task, while our RQ2 confirms the substantial benefits of the fine-tuning and the prompting component. The results show that both fine-tuning and effective prompting designs outperform LLMs with one single individual component alone.





\sectopic{Novelty \& Contributions.} 
\begin{enumerate}
\item We introduce a novel approach that utilizes Large Language Models (LLMs) for Text-to-Testcase generation based on fine-tuning with an effective prompting design. We are the first to investigate the performance of LLMs-based for the Text-to-Testcase generation task.

\item The ablation study confirms the substantial benefits of fine-tuning and prompt design for the Text-to-Testcase generation task.
\end{enumerate}

\sectopic{Open Science.}
To support open science, we publish the studied dataset, scripts (i.e., data processing and model fine-tuning), and experimental results in a GitHub repository.
(\url{https://github.com/LLMforTDD/LLMforTDD}).

\revise{R2.13}{\sectopic{Structure.} Section~\ref{sec:Background} discusses background and related work. 
Section~\ref{sec:Experimentaldesign} presents our experiment design. 
Section~\ref{sec:LLMforTDD} presents our approach. 
Section~\ref{sec:Results} presents the results. 
Section~\ref{sec:Discussion} presents the discussion.
Section~\ref{sec:Threats} discloses the threats to validity, and Section~\ref{sec:Conclusion} concludes the paper. }

\section{Background and Related Work}~\label{sec:Background}
In this section, we provide background knowledge of Test-Driven Development (TDD) and Large Language Models (LLMs) and discuss related work to motivate the paper.

\subsection{Test Driven Development (TDD)}

Test Driven Development (TDD)~\citep{beck2022test,astels2003test} is a software development practice that suggests writing test cases before implementing source code.
Starting with a comprehensive requirement, TDD follows an iterative process of providing a small unit test case that aligns with the requirement while developing a small chunk of source code to pass the test case at a time.
The practice aligns with the Agile Development Methodology that emphasizes fast continuous feedback loops and enhances collaboration between team members.
Recent research found that TDD can increase the understanding of requirements and produce a higher number of test cases than the traditional code-first method~\citep{mueller2002experiment,janzen2008does,erdogmus2005effectiveness}.
Despite the numerous benefits discussed in the literature, TDD is still manual, time-consuming, and labor-intensive, highlighting the need for automated Text-To-Testcase generation.

\subsection{Large Language Models (LLMs)}
A language model (LM) is a statistical model trained to predict the next word in a sequence~\citep{Jurafsky:20}, e.g., for the sequence ``Paris is the capital of'', it will predict with a very high likelihood that ``France'' is the next word in the sequence. 
Large language models are LMs with a substantially large number of weights and parameters and a complex training architecture trained to perform various downstream NLP tasks, e.g., text generation, question answering, and text classification~\citep{wei2022emergent}. 
The recent advancements in LLMs, most notably with chatbot platforms built on LLMs, e.g., OpenAI's ChatGPT, have led to several applications in software engineering (SE)~\citep{nguyen2023generative,hou2023large}. 
These include, among others, requirements engineering automation tasks~\citep{arora2023advancing}, code generation and completion~\citep{fan2023large}, software vulnerability~\citep{fu2023chatgpt}, test generation~\citep{yuan2023no}, and software maintenance~\citep{nguyen2023generative}.


The use of LLMs for generating test cases from text (text-to-testcase) is a relatively new application. 
It involves LLMs understanding a natural language (NL) description of the software method and then generating appropriate test cases that can be used to verify that the software meets its specifications. 
While there is interest in this area, it still remains underexplored due to the complexity of accurately interpreting descriptions and generating meaningful and comprehensive test cases. 
However, as LLMs become more advanced, their ability to perform such specialized tasks will likely improve, making text-to-testcase generation a promising area for future research and application in software engineering.

\sectopic{Prompting.}
In generative AI models (built on LLMs), the input is provided as a natural language instruction to elicit a desired response or output~\citep{brown2020language}. 
The input is known as a ``prompt''. How well a prompt is crafted determines the quality and relevance of the AI's output. There are different prompting strategies, and one typically needs to experiment with various prompts to elicit accurate responses or outputs from the models~\citep{nguyen2023generative}. 

\sectopic{Fine-tuning} 
In the context of Generative AI and ML refers to taking a pre-trained model and further training it on a specific, usually smaller, dataset to adapt it to a particular task or set of tasks. This process is prevalent in applications where a model, initially trained on a diverse and general dataset, needs to be specialized for more specific purposes. 
This specialized dataset is tailored to meet the particular needs or context of the intended application or task. This approach is beneficial as it leverages the extensive learning from the initial training phase, making the fine-tuning process more efficient than training a new model from scratch on the specialized data. Therefore, fine-tuning enhances the model's performance and accuracy in tasks requiring a more focused understanding or specialized knowledge~\citep{brown2020language}. 

\subsection{Related Work}
In this section, we discuss related work on automated test case generation, LLM in software engineering, and LLM for test case generation to highlight the significance of our paper.

\subsubsection{Automated Test Case Generation}
\label{sec:testgen}

Automated test case generation is an emerging technique that leverages neural networks and other deep learning architectures to automatically create test cases for software applications. 
This approach aims to enhance the efficiency and effectiveness of software testing by automating the generation of test inputs and expected outputs.
\revise{R1.2,R2.11}{In the past few years, researchers proposed automated test case generation approaches such as EvoSuite~\citep{fraser2011evosuite} and Randoop~\citep{pacheco2007randoop} based on different techniques, e.g., random-based approaches, search-based approaches and deep learning-based approaches.}

AthenaTest~\citep{tufano2020unit}, A3Test~\citep{alagarsamy2023a3test}, ChatUniTest~\citep{xie2023chatunitest} formulated a test case generation task as a neural machine translation problem, taking a Java method as input to generate test cases (Code$\rightarrow$Testcase).
However, these approaches often require a Java method as input, which is not applicable to Test-Driven Development practices that mandate writing test cases before writing code.
These limitations highlight the need for a text-to-testcase generation approach (Text$\rightarrow$Testcase) to support Test-Driven Development practices that are widely used in many software organizations~\citep{bissi2016effects}.

\subsubsection{LLMs for Test Case Generation}
Large Language Models (LLMs) have emerged as powerful tools in various fields of natural language understanding and generation, including software engineering~\citep{hou2023large}.
Notably, Hou~\ea~\citep{hou2023large} and Wang~\ea~\citep{wang2023software} found that LLMs have been widely used to support various software testing tasks.
For example, 
Schafer~\ea~\citep{schafer2023empirical} proposed TestPilot, a LLM-based test generation tool for JavaScript that automatically generates unit tests for all API functions in an npm package.
Siddiq~\ea~\citep{siddiq2023exploring} leveraged Large Language Models for test case generation.
Similarly, 
Yuan~\ea~\citep{yuan2023no} and Xie~\ea~\citep{xie2023chatunitest} leveraged ChatGPT for test case generation.
\revise{R2.3, R2.14}{Moradi Dakhel~\ea~\citep{dakhel2024effective} proposed an approach that utilizes pre-trained Large Language Models in conjunction with mutation testing to generate effective test cases.
Yang~\ea~\citep{yang2024empirical} conducted an empirical study to assess the performance of LLMs in unit test generation. 
Additionally, Hossain~\ea~\citep{hossain2024togll} introduced TOGLL, a framework for generating correct and robust test oracles using LLMs.
}

While LLMs have been widely used for test case generation, existing work only focuses on generating test cases based on a given Java method (Code$\rightarrow$Testcase), which does not support TDD, as TDD mandates writing test cases based on requirements before writing the actual code.

\subsubsection{Effectiveness of Test Case}
\revise{R2.4}{The effectiveness of test cases in revealing bugs is often measured using mutation testing~\citep{andrews2005mutation}, which evaluates a test suite's ability to detect small changes to the source code. This aspect is crucial as it reflects the practical utility of the test cases in a real-world testing scenario.
The literature consistently indicates that while code coverage is a valuable metric, it does not necessarily correlate strongly with the ability of tests to detect faults. 
Research has shown that traditional metrics, such as code coverage, do not necessarily correlate strongly with the effectiveness of test cases in identifying software defects.
Cai and Lyu~\citep{cai2005effect} suggest that higher coverage does not always equate to higher fault detection rates, especially under varying test profiles. 
Gopinath~\ea~\citep{gopinath2014code} examined that while coverage data is considered helpful, it is often insufficient for assessing the suite's effectiveness in catching critical bugs. 
Hemmati ~\citep{hemmati2015effective} provided a critical analysis of different code coverage criteria and their effectiveness in identifying bugs. 
In light of these findings, our research aims to bridge the gap by not only assessing test cases based on syntax correctness, code coverage, and requirement alignment, but also exploring their capability to detect bugs to understand the effectiveness of test case.}

\section{Experimental Design}
~\label{sec:Experimentaldesign}
In this section, we present the research questions and the experimental design. 
\subsection{Research Questions}

Our text-to-testcase generation approach involves two key components (fine-tuning and prompting).
Thus, our experiment aims to investigate the performance of our text-to-testcase generation approach when compared to other baseline models, conduct an ablation study to investigate the benefits of fine-tuning and prompting steps, and investigate the types of errors that occur in the incorrectly generated test cases.
Therefore, we formulate the following research questions.
\begin{enumerate}
    \item[\textbf{RQ1}]  \textbf{\rqone} 

    \item[\textbf{RQ2}]  \textbf{\rqtwo} 

    \item[\textbf{RQ3}]  \textbf{\rqthree} 

\end{enumerate}
\begin{table}
\centering
\caption{A Summary of the Evaluation Dataset for the LLM-based text-to-testcase generation models. 
}
\resizebox{\columnwidth}{!}{%
    \begin{tabular}{|l|l|l|l|l|}
    \hline
    \textbf{Project Abbr.} & \textbf{Domain} & \textbf{Version} & \textbf{\# LOC }& \textbf{\# Description} \\ \hline
    Commons-Lang  & Lang Utility & 3.12.0 & 56,787 & 1878 \\ \hline
    Commons-Cli  & Cli Cmd-line interface & 1.5.0 & 36,289 & 368 \\ \hline
    Commons-Csv  & Csv Data processing & 1.10.0 & 25,485 & 31 \\ \hline
    Gson  & Gson Serialization & 2.10.1 & 24,686 & 174 \\ \hline
    Jfreechart  & Chart Visualization & 1.5.4  & 152,686& 4599 \\ \hline
    \end{tabular}
}

\label{tab:projects}
\end{table}
\subsection{Evaluation Dataset}
\label{sec:Evaluation}

To mitigate the risk of data leakage between model fine-tuning and model evaluation, we decided to curate a new benchmark dataset for evaluating the LLMs for the text-to-testcase generation task. 
Table~\ref{tab:projects} provides a summary of the studied five open-source Java projects namely, Apache Commons Lang~\cite{Apache}, JFreeChart~\citep{JFreeChart}, Apache Common CLI~\citep{ApacheCommonsCLI}, Apache Common CSV~\citep{ApacheCommonsCSv}, Google Gson~\citep{googleGson}.
We selected these five projects due to the following reasons. 
We chose \emph{open-source software projects} to support the reproducibility and replicability of our work, allowing researchers and practitioners to reproduce experiments and validate our results.
We chose \emph{diverse domains and varying sizes of projects} to ensure that the results are not bound to specific contexts.

\subsection{Baseline Comparisons}

\revise{R1.7, R3.2, R3.9}{To fairly evaluate our proposed approach, we chose the other eight LLMs (i.e., GPT-3.5, Incoder, Starcoder, CodeT5, Bloom, CodeGemma, CodeLlama, and Gemini) as a baseline comparison.
We choose the following eight LLMs based on the various model size, types of model architectures, open/closed source, and fit in our GPU infrastructure. Table~\ref{tab:models} shows the analysis of the characteristics of the baseline models.}

\begin{table}[ht]
\centering
\small
\caption{Baseline Models.}
\label{tab:models}
\begin{tabular}{|l|l|l|}
\hline
\textbf{Model} & \textbf{Model Size} & \textbf{Open/Closed Source} \\ \hline
GPT-3.5        & NA                  & Closed Source               \\ \hline
Bloom          & 560M                & Open Source                 \\ \hline
CodeT5         & 770M                & Open Source                 \\ \hline
Gemini         & NA                  & Closed Source               \\ \hline
Starcoder      & 137M                & Open Source                 \\ \hline
Incoder        & 6.7M                & Open Source                 \\ \hline
CodeLlama      & 7B               & Open Source                 \\ \hline
CodeGemma      & 7B               & Open Source                 \\ \hline
\end{tabular}

\end{table}

To investigate the benefits of the fine-tuning and prompting steps for our approach, we conduct an ablation study by selectively removing or ablating one of the components and observing the impact on the overall performance of our approach.
Thus, we conducted a comprehensive experiment of the four variations (i.e., With and Without Finetune, and Basic and Advanced Prompt) on eight LLMs (i.e., GPT-3.5-turbo, Bloom, CodeT5, Gemini, Starcoder, Incoder, CodeLlama, CodeGemma), totaling 32 models.







\subsection{Hyper-Parameter Settings for Fine-Tuning.}
We fine-tune the two models, using the following settings.
We run our experiment on two NVIDIA GeForce RTX 3090 GPUs with 24 GB vRAM, an Intel(R) Core(TM) i9-9980XE CPU@ 3.00GHz with 36 core processors, and 64G RAM. 
The hyperparameter settings for fine-tuning are reported in Table~\ref{tab:hyperparameters}.

\begin{table}
\centering
\caption{Hyper-parameters and their values.}
\label{tab:hyperparameters}
\begin{tabular}{l|l}
\hline
\textbf{Hyperparameter}            & \textbf{Value} \\ \hline
Learning Rate                      & 2e-5   \\        
Warmup Steps                       & 1000   \\        
Weight Decay                       & 0.01   \\        
Batch Size                         & 2      \\        
Gradient Accumulation Steps        & 4        \\      
Learning Rate Scheduler Type       & inverse\_sqrt  \\ 
Epoch                              & 20             \\ \hline
\end{tabular}

\end{table}

\subsection{Evaluation Measures}
\begin{figure*}[t]
  \includegraphics[width=\linewidth]{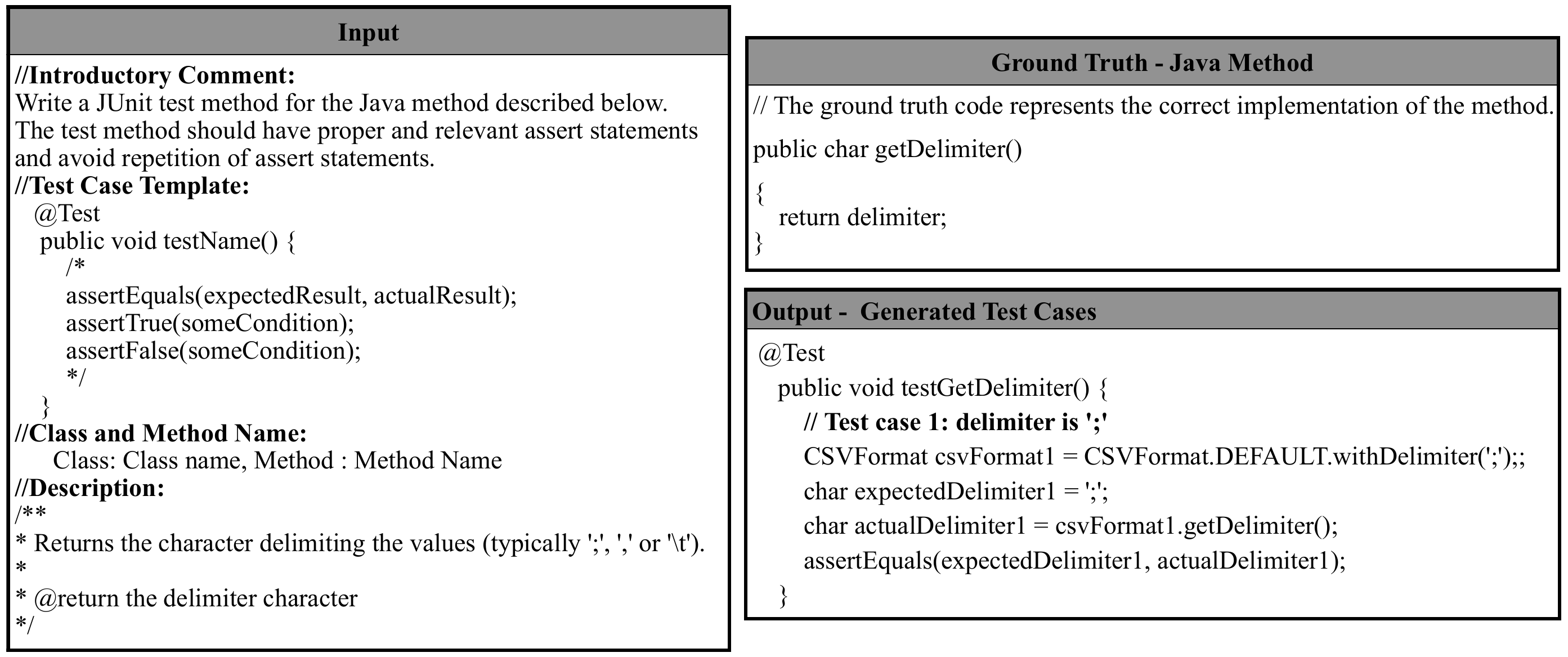}
  \caption{Model input and the generated test case and the ground-truth for evaluation.
  }
  \label{Input}
\end{figure*}

The LLMs take "Improved prompt and text" as input and the focal method as ground truth for evaluating the generated test cases (Figure~\ref{Input}).
We evaluate the quality of the test cases generated by LLMs based on the following four aspects:

\begin{itemize}
    \item \textbf{Syntax Correctness} measures the percentage of the generated test cases that are syntactically correct without any compiler errors, which is a prerequisite for the TDD process. 
    The given test case is considered syntactically correct if it can compile successfully. 
    \item \textbf{Requirements Alignment} measures the percentage of generated test cases that match the specified requirement. \revise{R2.10}{Only the syntactically correct test case are considered for the requirement alignment.}
    Since the ground truth code represents the correct implementation of the method requirement, any test cases that execute successfully against the ground truth code are considered correct (i.e., they correctly align with the given requirement and invoke the appropriate functions to be tested with the correct test inputs and expected outputs). 
    Therefore, we define that a test case is aligned with a given requirement if it can be executed with the ground truth code. 
    To assess this, we execute the combination of the ground-truth code and the generated test cases.
    \item \textbf{Code Coverage} measures the completeness of the generated test cases for a given description. 
    Ideally, in addition to aligning with the requirements, the test cases must be completed (i.e. executing all lines of the ground-truth code solution). Thus, we measure the line-level code coverage of the generated test cases using the Jacoco~\citep{Jacoco}.
\revise{R2.7,R2.4}{\item \textbf{Mutation Score} measures the effectiveness of the generated test cases. High-quality test cases should be capable of detecting bugs when the code undergoes slight modifications. Mutation testing introduces intentional faults (mutations) into the code to determine if the existing test cases can identify these faults. The mutation score is then calculated as the percentage of mutations detected (killed) out of the total number of mutations introduced. 
The mutation operators are used to systematically create variants of the program under test, each differing by only a small syntactic change. These operators include Arithmetic Operator Replacement (AOR), Logical Operator Replacement (LOR), Shift Operator Replacement (SOR), Relational Operator Replacement (COR), Increment Operator Replacement (ROR), Unary Operator Replacement (ORU), Literal Value Replacement (LVR), and Statement Deletion (STD). The test cases are executed on each mutant; if a test cases that passes on the original program fails on the mutant, then the mutant is killed. 
To analyse the mutation score, we use the PIT Test~\citep{coles2016pit} tool to measure the mutation score for our JUnit test cases. We applied all mutation operators: AOR, LOR, SOR, COR, ROR, ORU, LVR, STD. 
A high mutation score indicates that the generated test case is effective in detecting faults introduced by the mutations.}

\end{itemize}

\section{Enhancing Large Language Models for Text-To-Testcase
Generation}
~\label{sec:LLMforTDD}

\begin{figure*}[t]
    \centering
    \includegraphics[width=.99\textwidth]{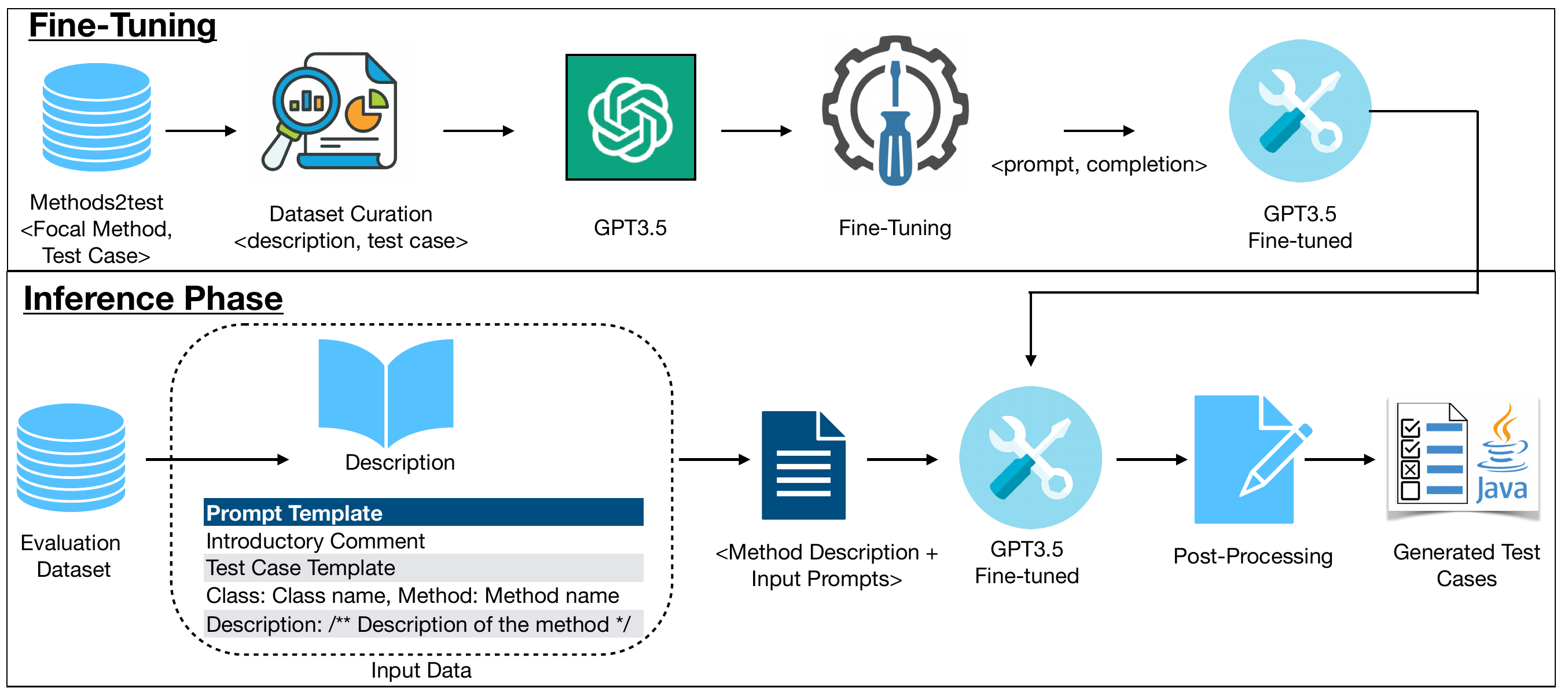}
  \caption{An overview of our approach (i.e.,  the prompting-based fine-tuning GPT-3.5-turbo model for the Text-To-Testcase generation task. 
  }
  \label{Overview}
\end{figure*}

\sectopic{Problem Formulation.} In this paper, we propose a text-to-testcase generation approach based on a fine-tuned LLM with an effective prompting design.
In particular, we formulate the text-to-testcase as a generation task using a fine-tuned LLM (\emph{GPT-3.5-turbo}). \revise{R2.12}{The model has been fine-tuned using a large dataset that was specifically mined and curated from various open-source Java projects.}
During the model fine-tuning, our LLM aims to learn the mapping between the method description (\emph{text}), which is the comments or documentation that describe the purpose, functionality, and usage of a method, and the corresponding test case (\emph{testcase}) which is a test method that includes a set of assert statements designed to verify whether the implemented method (\emph{method}) is aligned with the given requirement or not (i.e., functionally correct). 
Once fine-tuned, during the model inference phase, 
an unseen method description is used as input into the fine-tuned LLM to generate the corresponding test cases without requiring the implemented method (\emph{method}), which differs from most existing test case generation approaches.
Our approach consists of three phases: (1)~model fine-tuning, (2)~prompt design, and (3)~model inference.
Below, we describe the details of our approach (Figure~\ref{Overview}).

\subsection{Model Fine-tuning}
While LLMs demonstrate impressive capabilities in understanding and generating natural language~\citep{hou2023large}, they may lack the nuanced contextual understanding required for specialized applications, e.g., the text-to-testcase generation task.
The fine-tuning of LLMs is essential to enhance their performance and adaptability to our text-to-testcase generation task. 
Thus, it is important to have a representative and well-labeled dataset for the specific domain to achieve effective fine-tuning results.
However, the existing test case generation datasets are not suitable for our task.
For example, Tufano~\ea~\citep{tufano2022methods2test} introduced the Method2Test dataset, which is a pair of $<$\emph{Method,Testcase}$>$, but lacking the corresponding method description that is required for our approach.
Thus, we need to carefully curate the fine-tuning dataset.

\subsubsection{Dataset Curation}
\label{sec:DatasetCuration}

For the fine-tuning of the model, we started from
the Methods2Test dataset~\citep{tufano2022methods2test}.
This dataset consists of 780K pairs of Java focal methods and their corresponding test cases, collected from 91K open-source Java projects.
Since the method description is unavailable, we extended the Methods2Test dataset.
Starting from the 91K open-source Java projects, we cloned each project from the provided GitHub URL and performed the following steps:

\textbf{(Step 1) Collecting Code Structural Information}. 
The structural information of the code, such as method and class names are very important as we used to identify the relationship between the focal methods (i.e., methods to be tested) and their corresponding test methods.
To collect such information, we used the tree-sitter parser~\citep{tree-sitter} to analyze the structure of source code within the given project based on the detailed abstract syntax tree (AST) representation of source code.
With this step, we are able to identify the list of method and class names within the project.


\textbf{(Step 2) Identify Test Classes.} 
With the parser, we are able to identify the test classes (i.e., the classes specifically created to test the functionality of other classes in a program).
Thus, any class that contains at least one method with the \texttt{@Test} annotation is identified as a test class.
Such the \texttt{@Test} annotation in Java is normally used to signal that the annotated method should be executed as a test case. 

\textbf{(Step 3) Identify Focal Classes.}
For each test class in Step 2, we aim to identify the focal class associated with the given test class.
To do so, we employ the following two heuristics:

\textit{File Path Matching:}
Following the best practices of JUnit testing, they suggest placing focal classes and corresponding test classes in a mirrored directory structure.
For example, if a class is located at \texttt{src/main/java/Foo.java}, the corresponding JUnit test cases should be placed in the test class file \texttt{src/test/java/FooTest.java}.

\textit{File Name Matching:}
The name of a test class is usually composed of the name of the focal class, starting with a "Test" prefix or suffix. 
For example, the test case for the class \texttt{Foo.java} would probably be named as \texttt{FooTest.java}. 
Thus, we can identify the test classes and their corresponding focal classes with the heuristics of file path and file name matching.

\textbf{(Step 4) Identify Test Methods, Focal Methods.} 
We identify the test methods for each test class based on the \texttt{@Test} annotation.
Similar to Step 3, we apply the method name matching heuristic (i.e., removing \texttt{Test} prefix or suffix in the method name) to identify the corresponding focal method and its method description.

\revise{R1.4, R2.9}{\textbf{(Step 4) Identify Method Descriptions.} 
We utilized the Java parser to analyze the source code and identify the definitions of methods. 

\textit{Documentation Extraction:} We extracted the accompanying documentation comments, including Javadoc and inline comments, to gather details about each method. 
This process involved capturing the
Javadoc comments placed above methods, which provide a summary of their intended use and behavior.
We also extracted the inline comments embedded within the method bodies that offer additional context and explanations.
We then aggregated all relevant comments for each method to create a comprehensive description.}

With these steps, we construct the dataset \emph{$<$Text,Testcase,Method$>$} in the format of triplets. 
The dataset consists of 163k pairs of JUnit test cases and corresponding method description. 
\revise{R3.8}{Our dataset is derived from real-world systems and includes a variety of data characteristics that are typical of development environments, such as incomplete method descriptions and noisy input. 
Specifically, the dataset consists of method descriptions, comprising 109k pairs with 67\% of the total. Inline comments make up 13k pairs, representing 8\%, and combinations of method descriptions and inline comments account for 40k pairs with 25\% of the dataset.
This diversity ensures that our model is evaluated against different textual formats and challenges, including handling missing information. 
This comprehensive evaluation is crucial for ensuring the practical viability and robustness of your approach in real-world applications.
}
\revise{R1.5}{We split
the dataset into two sets, i.e., training set (60\% - 97,800 pairs), validation set (20\% - 32,600 pairs) and test set (20\% - 32,600 pairs). We used our validation data set with 32k pairs of \emph{$<$Text,Testcase$>$} for our finetuning process}
\textbf{We note that the methods will not be used for model fine-tuning like prior work, but are only used for evaluating test cases.}

\revise{R3.7}{Our dataset, used for fine-tuning the language model, comprises software methods with a broad spectrum of characteristics: method inline descriptions range from 16 to 1083 characters (average 39.54), method description vary between 39 and 1672 characters (average 303.52). This collection includes only the method type ``public'' drawn from various open-source projects to ensure a wide representation of programming styles and complexities. Additionally, the data set is designed to mitigate potential biases by including methods from multiple domains and programming paradigms, with the aim of providing a comprehensive foundation for model training.}

\subsubsection{Fine-Tuning}

When performing model fine-tuning, each training example generally consists of a single input (i.e., method description)  example and its associated output (i.e., test case) without giving detailed instructions or including multiple examples in the same prompt normally used for the base models (GPT-3.5).
Following the OpenAI's legacy fine-tuning best practices\footnote{https://platform.openai.com/docs/guides/legacy-fine-tuning}, each training example typically consists of a single input example and its associated output, which is typically constructed in the following format: \texttt{\{"input": "<Text>", "output": "<test case>"\}}.
We extract method descriptions and corresponding test cases into a list of prompts and completions for fine-tuning. 
We initiate a fine-tuning job using OpenAI's CreateFineTuneJob.create method, specifying the training data and model to fine-tune (GPT-3.5-turbo). 
We use ChatCompletion.create to generate test cases with the fine-tuned model, providing: The model engine ID from the fine-tuning job, the input is the method description prompt. 
The output is completion of the generated test case.The template includes the \texttt{\{"prompt": "<Text>", "completion": "<test case>"}. 
The input is a single input-output pair.
We used the curated dataset constructed in Section~\ref{sec:DatasetCuration} for model fine-tuning (the pairs of \texttt{<text, test case>}).
\revise{R3.8}{The cost for fine-tuning is \$0.0080 per 1,000 tokens, totaling \$68.7 for fine-tuning the GPT-3.5 model.}

Once completed, we use the fine-tuned model for generating test cases in the next step.

\begin{figure}[ht]
    \centering
    \includegraphics[width=0.55\linewidth]{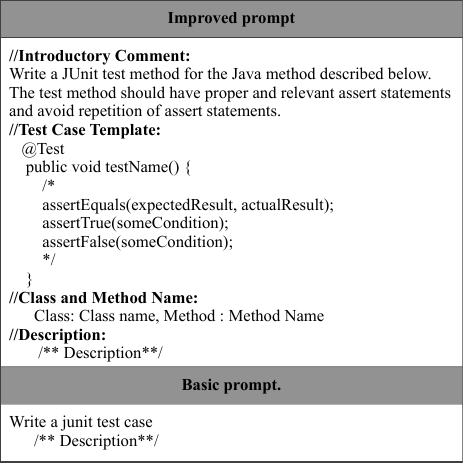} 
    \caption{Structure of our Improved prompt and Basic prompt. }
    \label{prompt}
\end{figure}

\subsection{Model Inference}
Once the tuning is complete, we use the fine-tuned model to perform inference tasks. To apply our approach using the fine-tuned model, we require input in the form of a prompt, with the textual description. We note that, in our study for inference, we developed an evaluation dataset (Section~\ref{sec:Evaluation}) with 7K test methods that span across five large-scale open-source software projects (i.e., Lang, Chart, Cli, Csv, Gson).
We used the java parser to identify the public methods and we extracted the method description of each public method for all the 5 projects.
The average cost associated with fine-tuning GPT-3.5 is \$0.0057 per test case and \$40.28 in total for all 5 evaluation projects.
\subsubsection{Test Case Generation}

We require the fine-tuned model and the prompt as input for test case generation. The model then generates a response containing the generated test cases. 

For test case generation, the most important step was to develop a prompt that works well, as the prompts provide context and guidance to the model. 
Without a (clear) prompt, the model would not know what information or response is expected. The process of formulating prompts for software engineering tasks is inherently subjective, and there are no established empirical guidelines for optimizing prompt effectiveness and robustness. We conducted several experiments involving various prompt combinations, ranging from basic to customized prompts, to address concerns associated with prompt optimization and devised our improved prompt. We present two prompts below. A basic prompt (our starting point) and an improved prompt (after several iterations). 
We explain each prompting technique below.

\sectopic{Basic prompt} 
For basic prompt, we first design the prompt template as presented in Figure~\ref{prompt} following the OpenAI guidelines~\footnote{\url{https://help.openai.com/en/articles/6654000-best-practices-for-promptengineering-with-openai-api}}~\footnote{\url{https://platform.openai.com/docs/guides/prompt-engineering/strategywrite-clear-instructions}}
to ensure
that the structure of the prompt is suitable for GPT-3.5. 
The prompt
template consists of the following components: an instruction and an
input (i.e., basic instruction and a text description).
Finally, we use the LLMs to generate test case from the
created prompts.
The structure of the basic prompt is as follows:

\texttt{"/* Write a junit test case
/**Description **/"}


\sectopic{Improved prompt} 
\revise{R3.1}{For Improved prompt, we first design the prompt template as presented in Figure~\ref{prompt}.
The prompting technique employed is One-shot prompting~\citep{brown2020language} which involves instructing the model with a single, detailed example or instruction to generate a specific output, such as a JUnit test method.
Similar to basic prompt, we follow the guidelines of OpenAI when designing the prompt template. 
The prompt template consists of the following components: demonstration
examples, instruction, and input (that is, the test case template
and target class and method name).
It is the iterated version based on multistep optimizations. 
We tested the prompt in the first step on a few samples from a small validation set (not from our curated or evaluation set to avoid data leakage), manually analyzed the generation performance and optimized the generation results by providing ChatGPT with a series of new prompts.

The approach to refining the prompt design for test case generation based on Test-Driven Development (TDD) principles~\cite{beck2003test} can be aligned with the three core TDD steps: Write a test, run it, and do it right. These principles were applied to disign the improved prompt.
We designed the prompt to instruct the LLM with a more specific context.
It emphasizes the importance of proper and relevant assertion statements, avoiding repetition, following a structured test case format, and adding context details for a more structured test case. A detailed breakdown of the prompt template components and their intended purpose is as follows:

\begin{itemize}
    \item 
\texttt{Introductory Comment:(Write a Test)} In TDD, the first step is to write a test that fails initially because the functionality it tests does not yet exist. This prompt design, sets the expectations for the model, defining the scope and focus of the test case it needs to generate. 
Instructs the model to focus on creating a test case, highlighting the need for relevant assert statements and avoiding repetition. 

\item 
\texttt{Test Case Template:}
Provides a skeleton for the test method structure, including the @Test annotation and placeholders for common assert statements.

\item 
\texttt{Class and Method Name:(Make it Run)}
After writing a failing test, the next step in TDD is to write enough code stubs to pass the test case. For the prompt design we use the "Class and Method Name" as our stub, directly guides the LLM to generate a test case relevant to the specific target class and method name for which the test is to be written, ensuring the generated test is relevant.

\item 
\texttt{Method Description:} 
The "Method Description" in the prompt plays a crucial role here, offering the LLM detailed context about what the method does, which enables it to generate more accurate and contextually relevant test assertions. Provides a description of the method, giving context to generate a test case.
\end{itemize}

}

Once the test cases are generated, we conduct a series of checks and corrections in post-processing to ensure the validity of these generated test cases.

\subsubsection{Post-Processing}

When extracting tests from LLM's response, syntactic errors may occur because the tokens limit is exceeded, resulting in truncated tests. 
\revise{R1.6}{To address this challenge, we introduce an automated verification approach in order to verify and correct the naming consistency and invalid test signatures.}
We thus perform a three-part post-processing
on the generated test cases.

\begin{enumerate}
    \item 
\textit{Verifying Test Signatures}: We inspect the beginning of each generated test case to ensure the presence of four specific keywords in sequence: @Test, public, void, and test[MethodName]. This sequence is crucial for the proper syntax of JUnit test methods.

\item 
\textit{Adding Missing Keywords:} If any of these keywords are missing in the initial sequence of the generated test method, we insert them accordingly. This step is essential to guarantee that the test methods can be executed successfully.

\item 
\textit{Correcting Incomplete Parentheses}: \revise{R1.8}{We noticed that some of the generated test cases contain incomplete parentheses, such as (or) is incomplete.} To address this issue, we perform corrections on these long sequence test cases, ensuring that all parentheses are correctly balanced.
\end{enumerate}

These measures ensure that the generated test methods align with the input method descriptions and adhere to the syntactical requirements for successful execution.

\begin{table*}[ht]
\caption{(RQ1)The experimental results of the ablation study
with eight different models ( Incoder, Starcoder, CodeT5, Bloom, CodeGemma, CodeLlama, Gemini and GPT-3.5).}
\label{ablationstudy}
\resizebox{\textwidth}{!}{%
\setlength{\tabcolsep}{0.5pt} 

\begin{tabular}{|c|l|rrrrrrrrrrrrrrrr|}
\hline
\multicolumn{1}{|l|}{} &  & \multicolumn{16}{c|}{\textbf{Evaluation Measures}} \\ \hline
\multicolumn{1}{|l|}{\textbf{Project}} & \textbf{Model} & \multicolumn{4}{c|}{\textbf{Syntax Correctness}} & \multicolumn{4}{c|}{\textbf{Requirement Alignment}} & \multicolumn{4}{c|}{\textbf{Code Coverage}} & \multicolumn{4}{c|}{\textbf{Mutation Score}} \\ \hline
\multirow{9}{*}{\textbf{Chart}} &  & \multicolumn{1}{l|}{FT+I.P} & \multicolumn{1}{l|}{FT+B.P} & \multicolumn{1}{l|}{NoFT+I.P} & \multicolumn{1}{l|}{NoFT+B.P} & \multicolumn{1}{l|}{FT+I.P} & \multicolumn{1}{l|}{FT+B.P} & \multicolumn{1}{l|}{NoFT+I.P} & \multicolumn{1}{l|}{NoFT+B.P} & \multicolumn{1}{l|}{FT+I.P} & \multicolumn{1}{l|}{FT+B.P} & \multicolumn{1}{l|}{NoFT+I.P} & \multicolumn{1}{l|}{NoFT+B.P} & \multicolumn{1}{l|}{FT+I.P} & \multicolumn{1}{l|}{FT+B.P} & \multicolumn{1}{l|}{NoFT+I.P} & \multicolumn{1}{l|}{NoFT+B.P} \\ \cline{2-18} 
 & Incoder & \multicolumn{1}{r|}{1.7} & \multicolumn{1}{r|}{1.14} & \multicolumn{1}{r|}{0.65} & \multicolumn{1}{r|}{0.18} & \multicolumn{1}{r|}{1.48} & \multicolumn{1}{r|}{0.99} & \multicolumn{1}{r|}{0.56} & \multicolumn{1}{r|}{0.15} & \multicolumn{1}{r|}{1.29} & \multicolumn{1}{r|}{0.86} & \multicolumn{1}{r|}{0.49} & \multicolumn{1}{r|}{0.13} & \multicolumn{1}{r|}{0.37} & \multicolumn{1}{r|}{0.25} & \multicolumn{1}{r|}{0.14} & 0.04 \\ \cline{2-18} 
 & Starcoder & \multicolumn{1}{r|}{2.4} & \multicolumn{1}{r|}{1.61} & \multicolumn{1}{r|}{0.92} & \multicolumn{1}{r|}{0.25} & \multicolumn{1}{r|}{2.09} & \multicolumn{1}{r|}{1.4} & \multicolumn{1}{r|}{0.8} & \multicolumn{1}{r|}{0.22} & \multicolumn{1}{r|}{1.82} & \multicolumn{1}{r|}{1.22} & \multicolumn{1}{r|}{0.7} & \multicolumn{1}{r|}{0.19} & \multicolumn{1}{r|}{0.53} & \multicolumn{1}{r|}{0.35} & \multicolumn{1}{r|}{0.2} & 0.06 \\ \cline{2-18} 
 & CodeT5 & \multicolumn{1}{r|}{8.32} & \multicolumn{1}{r|}{7.24} & \multicolumn{1}{r|}{4.13} & \multicolumn{1}{r|}{1.12} & \multicolumn{1}{r|}{3.56} & \multicolumn{1}{r|}{3.1} & \multicolumn{1}{r|}{1.77} & \multicolumn{1}{r|}{0.48} & \multicolumn{1}{r|}{3.03} & \multicolumn{1}{r|}{2.64} & \multicolumn{1}{r|}{1.5} & \multicolumn{1}{r|}{0.41} & \multicolumn{1}{r|}{0.88} & \multicolumn{1}{r|}{0.77} & \multicolumn{1}{r|}{0.44} & 0.12 \\ \cline{2-18} 
 & Bloom & \multicolumn{1}{r|}{26.02} & \multicolumn{1}{r|}{22.64} & \multicolumn{1}{r|}{12.9} & \multicolumn{1}{r|}{3.48} & \multicolumn{1}{r|}{22.94} & \multicolumn{1}{r|}{19.96} & \multicolumn{1}{r|}{11.38} & \multicolumn{1}{r|}{3.07} & \multicolumn{1}{r|}{15.93} & \multicolumn{1}{r|}{13.86} & \multicolumn{1}{r|}{7.9} & \multicolumn{1}{r|}{2.13} & \multicolumn{1}{r|}{4.62} & \multicolumn{1}{r|}{4.02} & \multicolumn{1}{r|}{2.29} & 0.62 \\ \cline{2-18} 
 & CodeGemma & \multicolumn{1}{r|}{49.7} & \multicolumn{1}{r|}{33.3} & \multicolumn{1}{r|}{18.98} & \multicolumn{1}{r|}{5.12} & \multicolumn{1}{r|}{43.24} & \multicolumn{1}{r|}{28.97} & \multicolumn{1}{r|}{16.51} & \multicolumn{1}{r|}{4.46} & \multicolumn{1}{r|}{37.62} & \multicolumn{1}{r|}{25.21} & \multicolumn{1}{r|}{14.37} & \multicolumn{1}{r|}{3.88} & \multicolumn{1}{r|}{10.91} & \multicolumn{1}{r|}{7.31} & \multicolumn{1}{r|}{4.17} & 1.13 \\ \cline{2-18} 
 & CodeLlama & \multicolumn{1}{r|}{56.79} & \multicolumn{1}{r|}{38.05} & \multicolumn{1}{r|}{21.69} & \multicolumn{1}{r|}{5.86} & \multicolumn{1}{r|}{49.41} & \multicolumn{1}{r|}{33.1} & \multicolumn{1}{r|}{18.87} & \multicolumn{1}{r|}{5.09} & \multicolumn{1}{r|}{42.99} & \multicolumn{1}{r|}{28.8} & \multicolumn{1}{r|}{16.42} & \multicolumn{1}{r|}{4.43} & \multicolumn{1}{r|}{12.47} & \multicolumn{1}{r|}{8.35} & \multicolumn{1}{r|}{4.76} & 1.28 \\ \cline{2-18} 
 & Gemini & \multicolumn{1}{r|}{74.8} & \multicolumn{1}{r|}{50.12} & \multicolumn{1}{r|}{28.57} & \multicolumn{1}{r|}{7.71} & \multicolumn{1}{r|}{65.08} & \multicolumn{1}{r|}{43.6} & \multicolumn{1}{r|}{24.85} & \multicolumn{1}{r|}{6.71} & \multicolumn{1}{r|}{56.62} & \multicolumn{1}{r|}{37.94} & \multicolumn{1}{r|}{21.63} & \multicolumn{1}{r|}{5.84} & \multicolumn{1}{r|}{16.42} & \multicolumn{1}{r|}{11} & \multicolumn{1}{r|}{6.27} & 1.69 \\ \cline{2-18} 
 & \textbf{ GPT-3.5} & \multicolumn{1}{r|}{\textbf{82.4}} & \multicolumn{1}{r|}{46.09} & \multicolumn{1}{r|}{26.48} & \multicolumn{1}{r|}{17.4} & \multicolumn{1}{r|}{\textbf{73.7}} & \multicolumn{1}{r|}{35.17} & \multicolumn{1}{r|}{25.19} & \multicolumn{1}{r|}{13.91} & \multicolumn{1}{r|}{\textbf{69.6}} & \multicolumn{1}{r|}{31.43} & \multicolumn{1}{r|}{21.3} & \multicolumn{1}{r|}{12.91} & \multicolumn{1}{r|}{\textbf{20.18}} & \multicolumn{1}{r|}{9.11} & \multicolumn{1}{r|}{6.18} & 3.74
 \\ \hline
\multirow{8}{*}{\textbf{Cli}} & Incoder & \multicolumn{1}{r|}{1.67} & \multicolumn{1}{r|}{1.12} & \multicolumn{1}{r|}{0.64} & \multicolumn{1}{r|}{0.17} & \multicolumn{1}{r|}{1.45} & \multicolumn{1}{r|}{0.97} & \multicolumn{1}{r|}{0.55} & \multicolumn{1}{r|}{0.15} & \multicolumn{1}{r|}{1.26} & \multicolumn{1}{r|}{0.84} & \multicolumn{1}{r|}{0.48} & \multicolumn{1}{r|}{0.13} & \multicolumn{1}{r|}{0.37} & \multicolumn{1}{r|}{0.24} & \multicolumn{1}{r|}{0.14} & 0.04 \\ \cline{2-18} 
 & Starcoder & \multicolumn{1}{r|}{2.35} & \multicolumn{1}{r|}{1.57} & \multicolumn{1}{r|}{0.89} & \multicolumn{1}{r|}{0.24} & \multicolumn{1}{r|}{2.04} & \multicolumn{1}{r|}{1.37} & \multicolumn{1}{r|}{0.78} & \multicolumn{1}{r|}{0.21} & \multicolumn{1}{r|}{1.77} & \multicolumn{1}{r|}{1.19} & \multicolumn{1}{r|}{0.68} & \multicolumn{1}{r|}{0.18} & \multicolumn{1}{r|}{0.51} & \multicolumn{1}{r|}{0.35} & \multicolumn{1}{r|}{0.2} & 0.05 \\ \cline{2-18} 
 & CodeT5 & \multicolumn{1}{r|}{7.3} & \multicolumn{1}{r|}{6.35} & \multicolumn{1}{r|}{3.62} & \multicolumn{1}{r|}{0.98} & \multicolumn{1}{r|}{1.9} & \multicolumn{1}{r|}{1.65} & \multicolumn{1}{r|}{0.94} & \multicolumn{1}{r|}{0.25} & \multicolumn{1}{r|}{1.74} & \multicolumn{1}{r|}{1.51} & \multicolumn{1}{r|}{0.86} & \multicolumn{1}{r|}{0.23} & \multicolumn{1}{r|}{0.5} & \multicolumn{1}{r|}{0.44} & \multicolumn{1}{r|}{0.25} & 0.07 \\ \cline{2-18} 
 & Bloom & \multicolumn{1}{r|}{16.3} & \multicolumn{1}{r|}{14.18} & \multicolumn{1}{r|}{8.08} & \multicolumn{1}{r|}{2.18} & \multicolumn{1}{r|}{11.89} & \multicolumn{1}{r|}{10.34} & \multicolumn{1}{r|}{5.89} & \multicolumn{1}{r|}{1.59} & \multicolumn{1}{r|}{10.44} & \multicolumn{1}{r|}{9.08} & \multicolumn{1}{r|}{5.18} & \multicolumn{1}{r|}{1.4} & \multicolumn{1}{r|}{3.03} & \multicolumn{1}{r|}{2.63} & \multicolumn{1}{r|}{1.5} & 0.41 \\ \cline{2-18} 
 & CodeGemma & \multicolumn{1}{r|}{48.71} & \multicolumn{1}{r|}{32.64} & \multicolumn{1}{r|}{18.6} & \multicolumn{1}{r|}{5.02} & \multicolumn{1}{r|}{42.38} & \multicolumn{1}{r|}{28.39} & \multicolumn{1}{r|}{16.18} & \multicolumn{1}{r|}{4.37} & \multicolumn{1}{r|}{36.87} & \multicolumn{1}{r|}{24.7} & \multicolumn{1}{r|}{14.08} & \multicolumn{1}{r|}{3.8} & \multicolumn{1}{r|}{10.69} & \multicolumn{1}{r|}{7.16} & \multicolumn{1}{r|}{4.08} & 1.1 \\ \cline{2-18} 
 & CodeLlama & \multicolumn{1}{r|}{55.65} & \multicolumn{1}{r|}{37.29} & \multicolumn{1}{r|}{21.26} & \multicolumn{1}{r|}{5.74} & \multicolumn{1}{r|}{48.42} & \multicolumn{1}{r|}{32.44} & \multicolumn{1}{r|}{18.49} & \multicolumn{1}{r|}{4.99} & \multicolumn{1}{r|}{42.13} & \multicolumn{1}{r|}{28.23} & \multicolumn{1}{r|}{16.09} & \multicolumn{1}{r|}{4.34} & \multicolumn{1}{r|}{12.22} & \multicolumn{1}{r|}{8.19} & \multicolumn{1}{r|}{4.67} & 1.26 \\ \cline{2-18} 
 & Gemini & \multicolumn{1}{r|}{67} & \multicolumn{1}{r|}{51.59} & \multicolumn{1}{r|}{29.41} & \multicolumn{1}{r|}{7.94} & \multicolumn{1}{r|}{66.99} & \multicolumn{1}{r|}{44.88} & \multicolumn{1}{r|}{25.58} & \multicolumn{1}{r|}{6.91} & \multicolumn{1}{r|}{58.28} & \multicolumn{1}{r|}{39.05} & \multicolumn{1}{r|}{22.26} & \multicolumn{1}{r|}{6.01} & \multicolumn{1}{r|}{16.9} & \multicolumn{1}{r|}{11.32} & \multicolumn{1}{r|}{6.46} & 1.74 \\ \cline{2-18} 
& \textbf{GPT-3.5} & \multicolumn{1}{r|}{\textbf{73.7}} & \multicolumn{1}{r|}{37.8} & \multicolumn{1}{r|}{26.08} & \multicolumn{1}{r|}{11.08} & \multicolumn{1}{r|}{\textbf{62.5}} & \multicolumn{1}{r|}{26.05} & \multicolumn{1}{r|}{15.05} & \multicolumn{1}{r|}{9.81} & \multicolumn{1}{r|}{\textbf{59.5}} & \multicolumn{1}{r|}{23.04} & \multicolumn{1}{r|}{12.78} & \multicolumn{1}{r|}{8.09} & \multicolumn{1}{r|}{\textbf{17.26}} & \multicolumn{1}{r|}{6.68} & \multicolumn{1}{r|}{3.71} & 2.35 \\ \hline
\multirow{8}{*}{\textbf{Csv}} & Incoder & \multicolumn{1}{r|}{1.09} & \multicolumn{1}{r|}{0.73} & \multicolumn{1}{r|}{0.42} & \multicolumn{1}{r|}{0.11} & \multicolumn{1}{r|}{0.95} & \multicolumn{1}{r|}{0.64} & \multicolumn{1}{r|}{0.36} & \multicolumn{1}{r|}{0.1} & \multicolumn{1}{r|}{0.83} & \multicolumn{1}{r|}{0.56} & \multicolumn{1}{r|}{0.32} & \multicolumn{1}{r|}{0.09} & \multicolumn{1}{r|}{0.24} & \multicolumn{1}{r|}{0.16} & \multicolumn{1}{r|}{0.09} & 0.03 \\ \cline{2-18} 
 & Starcoder & \multicolumn{1}{r|}{2.67} & \multicolumn{1}{r|}{1.79} & \multicolumn{1}{r|}{1.02} & \multicolumn{1}{r|}{0.28} & \multicolumn{1}{r|}{2.32} & \multicolumn{1}{r|}{1.55} & \multicolumn{1}{r|}{0.88} & \multicolumn{1}{r|}{0.24} & \multicolumn{1}{r|}{2.02} & \multicolumn{1}{r|}{1.35} & \multicolumn{1}{r|}{0.77} & \multicolumn{1}{r|}{0.21} & \multicolumn{1}{r|}{0.59} & \multicolumn{1}{r|}{0.39} & \multicolumn{1}{r|}{0.22} & 0.06 \\ \cline{2-18} 
 & CodeT5 & \multicolumn{1}{r|}{7.05} & \multicolumn{1}{r|}{6.13} & \multicolumn{1}{r|}{3.49} & \multicolumn{1}{r|}{0.94} & \multicolumn{1}{r|}{2.07} & \multicolumn{1}{r|}{1.8} & \multicolumn{1}{r|}{1.03} & \multicolumn{1}{r|}{0.28} & \multicolumn{1}{r|}{1.78} & \multicolumn{1}{r|}{1.55} & \multicolumn{1}{r|}{0.88} & \multicolumn{1}{r|}{0.24} & \multicolumn{1}{r|}{0.52} & \multicolumn{1}{r|}{0.45} & \multicolumn{1}{r|}{0.26} & 0.07 \\ \cline{2-18} 
 & Bloom & \multicolumn{1}{r|}{17.57} & \multicolumn{1}{r|}{15.29} & \multicolumn{1}{r|}{8.72} & \multicolumn{1}{r|}{2.35} & \multicolumn{1}{r|}{12.07} & \multicolumn{1}{r|}{10.5} & \multicolumn{1}{r|}{5.99} & \multicolumn{1}{r|}{1.62} & \multicolumn{1}{r|}{11.98} & \multicolumn{1}{r|}{10.42} & \multicolumn{1}{r|}{5.94} & \multicolumn{1}{r|}{1.6} & \multicolumn{1}{r|}{3.47} & \multicolumn{1}{r|}{3.02} & \multicolumn{1}{r|}{1.72} & 0.46 \\ \cline{2-18} 
 & CodeGemma & \multicolumn{1}{r|}{44.78} & \multicolumn{1}{r|}{30} & \multicolumn{1}{r|}{17.1} & \multicolumn{1}{r|}{4.62} & \multicolumn{1}{r|}{38.96} & \multicolumn{1}{r|}{26.1} & \multicolumn{1}{r|}{14.88} & \multicolumn{1}{r|}{4.02} & \multicolumn{1}{r|}{33.9} & \multicolumn{1}{r|}{22.71} & \multicolumn{1}{r|}{12.94} & \multicolumn{1}{r|}{3.49} & \multicolumn{1}{r|}{9.83} & \multicolumn{1}{r|}{6.59} & \multicolumn{1}{r|}{3.75} & 1.01 \\ \cline{2-18} 
 & CodeLlama & \multicolumn{1}{r|}{50.5} & \multicolumn{1}{r|}{33.84} & \multicolumn{1}{r|}{19.29} & \multicolumn{1}{r|}{5.21} & \multicolumn{1}{r|}{43.94} & \multicolumn{1}{r|}{29.44} & \multicolumn{1}{r|}{16.78} & \multicolumn{1}{r|}{4.53} & \multicolumn{1}{r|}{38.23} & \multicolumn{1}{r|}{25.61} & \multicolumn{1}{r|}{14.6} & \multicolumn{1}{r|}{3.94} & \multicolumn{1}{r|}{11.09} & \multicolumn{1}{r|}{7.43} & \multicolumn{1}{r|}{4.23} & 1.14 \\ \cline{2-18} 
 & Gemini & \multicolumn{1}{r|}{71} & \multicolumn{1}{r|}{47.57} & \multicolumn{1}{r|}{27.11} & \multicolumn{1}{r|}{7.32} & \multicolumn{1}{r|}{61.77} & \multicolumn{1}{r|}{41.39} & \multicolumn{1}{r|}{23.59} & \multicolumn{1}{r|}{6.37} & \multicolumn{1}{r|}{53.74} & \multicolumn{1}{r|}{36.01} & \multicolumn{1}{r|}{20.53} & \multicolumn{1}{r|}{5.54} & \multicolumn{1}{r|}{15.58} & \multicolumn{1}{r|}{10.44} & \multicolumn{1}{r|}{5.95} & 1.61 \\ \cline{2-18} 
 & \textbf{GPT-3.5 } & \multicolumn{1}{r|}{\textbf{77.8}} & \multicolumn{1}{r|}{38.56} & \multicolumn{1}{r|}{28.78} & \multicolumn{1}{r|}{10.7} & \multicolumn{1}{r|}{\textbf{64.1}} & \multicolumn{1}{r|}{27.06} & \multicolumn{1}{r|}{18.01} & \multicolumn{1}{r|}{10.14} & \multicolumn{1}{r|}{\textbf{61.1}} & \multicolumn{1}{r|}{24.16} & \multicolumn{1}{r|}{14.16} & \multicolumn{1}{r|}{9.92} & \multicolumn{1}{r|}{\textbf{17.72}} & \multicolumn{1}{r|}{7.01} & \multicolumn{1}{r|}{4.11} & 2.88 \\ \hline
\multirow{8}{*}{\textbf{Gson}} & Incoder & \multicolumn{1}{r|}{1.99} & \multicolumn{1}{r|}{1.33} & \multicolumn{1}{r|}{0.76} & \multicolumn{1}{r|}{0.21} & \multicolumn{1}{r|}{1.73} & \multicolumn{1}{r|}{1.16} & \multicolumn{1}{r|}{0.66} & \multicolumn{1}{r|}{0.18} & \multicolumn{1}{r|}{1.51} & \multicolumn{1}{r|}{1.01} & \multicolumn{1}{r|}{0.58} & \multicolumn{1}{r|}{0.16} & \multicolumn{1}{r|}{0.44} & \multicolumn{1}{r|}{0.29} & \multicolumn{1}{r|}{0.17} & 0.05 \\ \cline{2-18} 
 & Starcoder & \multicolumn{1}{r|}{3.23} & \multicolumn{1}{r|}{2.16} & \multicolumn{1}{r|}{1.23} & \multicolumn{1}{r|}{0.33} & \multicolumn{1}{r|}{2.81} & \multicolumn{1}{r|}{1.88} & \multicolumn{1}{r|}{1.07} & \multicolumn{1}{r|}{0.29} & \multicolumn{1}{r|}{2.44} & \multicolumn{1}{r|}{1.63} & \multicolumn{1}{r|}{0.93} & \multicolumn{1}{r|}{0.25} & \multicolumn{1}{r|}{0.71} & \multicolumn{1}{r|}{0.47} & \multicolumn{1}{r|}{0.27} & 0.07 \\ \cline{2-18} 
 & CodeT5 & \multicolumn{1}{r|}{7.31} & \multicolumn{1}{r|}{6.36} & \multicolumn{1}{r|}{3.63} & \multicolumn{1}{r|}{0.98} & \multicolumn{1}{r|}{3.14} & \multicolumn{1}{r|}{2.73} & \multicolumn{1}{r|}{1.56} & \multicolumn{1}{r|}{0.42} & \multicolumn{1}{r|}{2.67} & \multicolumn{1}{r|}{2.32} & \multicolumn{1}{r|}{1.32} & \multicolumn{1}{r|}{0.36} & \multicolumn{1}{r|}{0.77} & \multicolumn{1}{r|}{0.67} & \multicolumn{1}{r|}{0.38} & 0.1 \\ \cline{2-18} 
 & Bloom & \multicolumn{1}{r|}{19.73} & \multicolumn{1}{r|}{17.17} & \multicolumn{1}{r|}{9.79} & \multicolumn{1}{r|}{2.64} & \multicolumn{1}{r|}{15.14} & \multicolumn{1}{r|}{13.17} & \multicolumn{1}{r|}{7.51} & \multicolumn{1}{r|}{2.03} & \multicolumn{1}{r|}{13.67} & \multicolumn{1}{r|}{11.89} & \multicolumn{1}{r|}{6.78} & \multicolumn{1}{r|}{1.83} & \multicolumn{1}{r|}{3.96} & \multicolumn{1}{r|}{3.45} & \multicolumn{1}{r|}{1.97} & 0.53 \\ \cline{2-18} 
 & CodeGemma & \multicolumn{1}{r|}{49.08} & \multicolumn{1}{r|}{32.88} & \multicolumn{1}{r|}{18.74} & \multicolumn{1}{r|}{5.06} & \multicolumn{1}{r|}{42.7} & \multicolumn{1}{r|}{28.61} & \multicolumn{1}{r|}{16.31} & \multicolumn{1}{r|}{4.4} & \multicolumn{1}{r|}{37.15} & \multicolumn{1}{r|}{24.89} & \multicolumn{1}{r|}{14.19} & \multicolumn{1}{r|}{3.83} & \multicolumn{1}{r|}{10.77} & \multicolumn{1}{r|}{7.22} & \multicolumn{1}{r|}{4.12} & 1.11 \\ \cline{2-18} 
 & CodeLlama & \multicolumn{1}{r|}{55.89} & \multicolumn{1}{r|}{37.45} & \multicolumn{1}{r|}{21.35} & \multicolumn{1}{r|}{5.76} & \multicolumn{1}{r|}{48.62} & \multicolumn{1}{r|}{32.58} & \multicolumn{1}{r|}{18.57} & \multicolumn{1}{r|}{5.01} & \multicolumn{1}{r|}{42.3} & \multicolumn{1}{r|}{28.34} & \multicolumn{1}{r|}{16.15} & \multicolumn{1}{r|}{4.36} & \multicolumn{1}{r|}{12.27} & \multicolumn{1}{r|}{8.22} & \multicolumn{1}{r|}{4.68} & 1.26 \\ \cline{2-18} 
 & Gemini & \multicolumn{1}{r|}{70.9} & \multicolumn{1}{r|}{47.5} & \multicolumn{1}{r|}{27.08} & \multicolumn{1}{r|}{7.31} & \multicolumn{1}{r|}{61.68} & \multicolumn{1}{r|}{41.33} & \multicolumn{1}{r|}{23.56} & \multicolumn{1}{r|}{6.36} & \multicolumn{1}{r|}{53.66} & \multicolumn{1}{r|}{35.95} & \multicolumn{1}{r|}{20.49} & \multicolumn{1}{r|}{5.53} & \multicolumn{1}{r|}{15.56} & \multicolumn{1}{r|}{10.43} & \multicolumn{1}{r|}{5.94} & 1.6 \\ \cline{2-18} 
 & \textbf{GPT-3.5} & \multicolumn{1}{r|}{\textbf{80.5}} & \multicolumn{1}{r|}{39.89} & \multicolumn{1}{r|}{29.84} & \multicolumn{1}{r|}{9.91} & \multicolumn{1}{r|}{\textbf{70.9}} & \multicolumn{1}{r|}{28.1} & \multicolumn{1}{r|}{20.01} & \multicolumn{1}{r|}{9.82} & \multicolumn{1}{r|}{\textbf{70.8}} & \multicolumn{1}{r|}{24.49} & \multicolumn{1}{r|}{14.89} & \multicolumn{1}{r|}{8.72} & \multicolumn{1}{r|}{\textbf{20.53}} & \multicolumn{1}{r|}{7.1} & \multicolumn{1}{r|}{4.32} & 2.53
\\ \hline
\multirow{8}{*}{\textbf{Lang}} & Incoder & \multicolumn{1}{r|}{1.06} & \multicolumn{1}{r|}{0.71} & \multicolumn{1}{r|}{0.4} & \multicolumn{1}{r|}{0.11} & \multicolumn{1}{r|}{0.92} & \multicolumn{1}{r|}{0.62} & \multicolumn{1}{r|}{0.35} & \multicolumn{1}{r|}{0.09} & \multicolumn{1}{r|}{0.8} & \multicolumn{1}{r|}{0.54} & \multicolumn{1}{r|}{0.31} & \multicolumn{1}{r|}{0.08} & \multicolumn{1}{r|}{0.23} & \multicolumn{1}{r|}{0.16} & \multicolumn{1}{r|}{0.09} & 0.02 \\ \cline{2-18} 
 & Starcoder & \multicolumn{1}{r|}{2.98} & \multicolumn{1}{r|}{2} & \multicolumn{1}{r|}{1.14} & \multicolumn{1}{r|}{0.31} & \multicolumn{1}{r|}{2.59} & \multicolumn{1}{r|}{1.74} & \multicolumn{1}{r|}{0.99} & \multicolumn{1}{r|}{0.27} & \multicolumn{1}{r|}{2.25} & \multicolumn{1}{r|}{1.51} & \multicolumn{1}{r|}{0.86} & \multicolumn{1}{r|}{0.23} & \multicolumn{1}{r|}{0.65} & \multicolumn{1}{r|}{0.44} & \multicolumn{1}{r|}{0.25} & 0.07 \\ \cline{2-18} 
 & CodeT5 & \multicolumn{1}{r|}{6.9} & \multicolumn{1}{r|}{6} & \multicolumn{1}{r|}{3.42} & \multicolumn{1}{r|}{0.92} & \multicolumn{1}{r|}{1.56} & \multicolumn{1}{r|}{1.36} & \multicolumn{1}{r|}{0.78} & \multicolumn{1}{r|}{0.21} & \multicolumn{1}{r|}{1.05} & \multicolumn{1}{r|}{0.91} & \multicolumn{1}{r|}{0.52} & \multicolumn{1}{r|}{0.14} & \multicolumn{1}{r|}{0.3} & \multicolumn{1}{r|}{0.26} & \multicolumn{1}{r|}{0.15} & 0.04 \\ \cline{2-18} 
 & Bloom & \multicolumn{1}{r|}{23.99} & \multicolumn{1}{r|}{20.87} & \multicolumn{1}{r|}{11.9} & \multicolumn{1}{r|}{3.21} & \multicolumn{1}{r|}{19.86} & \multicolumn{1}{r|}{17.28} & \multicolumn{1}{r|}{9.85} & \multicolumn{1}{r|}{2.66} & \multicolumn{1}{r|}{13.05} & \multicolumn{1}{r|}{11.35} & \multicolumn{1}{r|}{6.47} & \multicolumn{1}{r|}{1.75} & \multicolumn{1}{r|}{3.78} & \multicolumn{1}{r|}{3.29} & \multicolumn{1}{r|}{1.88} & 0.51 \\ \cline{2-18} 
 & CodeGemma & \multicolumn{1}{r|}{41.45} & \multicolumn{1}{r|}{27.77} & \multicolumn{1}{r|}{15.83} & \multicolumn{1}{r|}{4.27} & \multicolumn{1}{r|}{36.06} & \multicolumn{1}{r|}{24.16} & \multicolumn{1}{r|}{13.77} & \multicolumn{1}{r|}{3.72} & \multicolumn{1}{r|}{31.37} & \multicolumn{1}{r|}{21.02} & \multicolumn{1}{r|}{11.98} & \multicolumn{1}{r|}{3.23} & \multicolumn{1}{r|}{9.1} & \multicolumn{1}{r|}{6.1} & \multicolumn{1}{r|}{3.47} & 0.94 \\ \cline{2-18} 
 & CodeLlama & \multicolumn{1}{r|}{49.01} & \multicolumn{1}{r|}{32.84} & \multicolumn{1}{r|}{18.72} & \multicolumn{1}{r|}{5.05} & \multicolumn{1}{r|}{42.64} & \multicolumn{1}{r|}{28.57} & \multicolumn{1}{r|}{16.28} & \multicolumn{1}{r|}{4.4} & \multicolumn{1}{r|}{37.1} & \multicolumn{1}{r|}{24.86} & \multicolumn{1}{r|}{14.17} & \multicolumn{1}{r|}{3.83} & \multicolumn{1}{r|}{10.76} & \multicolumn{1}{r|}{7.21} & \multicolumn{1}{r|}{4.11} & 1.11 \\ \cline{2-18} 
 & Gemini & \multicolumn{1}{r|}{68.05} & \multicolumn{1}{r|}{45.59} & \multicolumn{1}{r|}{25.99} & \multicolumn{1}{r|}{7.02} & \multicolumn{1}{r|}{59.2} & \multicolumn{1}{r|}{39.66} & \multicolumn{1}{r|}{22.61} & \multicolumn{1}{r|}{6.1} & \multicolumn{1}{r|}{51.5} & \multicolumn{1}{r|}{34.51} & \multicolumn{1}{r|}{19.67} & \multicolumn{1}{r|}{5.31} & \multicolumn{1}{r|}{14.94} & \multicolumn{1}{r|}{10.01} & \multicolumn{1}{r|}{5.7} & 1.54 \\ \cline{2-18} 
& \textbf{GPT-3.5} & \multicolumn{1}{r|}{\textbf{76.9}} & \multicolumn{1}{r|}{45.8} & \multicolumn{1}{r|}{33.8} & \multicolumn{1}{r|}{15.74} & \multicolumn{1}{r|}{\textbf{63.9}} & \multicolumn{1}{r|}{32.4} & \multicolumn{1}{r|}{22.47} & \multicolumn{1}{r|}{11.87} & \multicolumn{1}{r|}{\textbf{62.9}} & \multicolumn{1}{r|}{27.4} & \multicolumn{1}{r|}{17.98} & \multicolumn{1}{r|}{10.95} & \multicolumn{1}{r|}{\textbf{18.24}} & \multicolumn{1}{r|}{7.95} & \multicolumn{1}{r|}{5.21} & 3.18
\\ \hline
\end{tabular}%
}

\end{table*}


\section{Experimental Results }~\label{sec:Results}
In this section, we present the results with respect to our three research questions. 
\subsection*{\textbf{RQ1: \rqone}}

\textbf{Results.} \textbf{Our proposed approach (i.e., the fine-tuned GPT-3.5-turbo with an effective prompting design) outperforms all other LLMs (basic GPT-3.5-turbo, Incoder, Starcoder, CodeT5, Bloom, CodeGemma, CodeLlama, and Gemini.}
Table~\ref{ablationstudy} presents the results of our approach and the baseline models for a text-to-testcase generation task. 

The application of fine-tuning and prompting methodologies did not consistently enhance the performance of various Large Language Models (LLMs) such as Incoder, Starcoder, CodeT5, these models achieve a very low-performing performance (e.g., less than 1-10\% syntax correctness of the generated test cases), meaning that they are not capable to be used in practices
in the domain of test case generation using TDD. 
The other models such as Bloom, CodeGemma, CodeLama, Gemini, and GPT-3.5 models exhibit diverse efficacy across multiple evaluation metrics including syntax correctness, requirement alignment, code coverage, and mutation scores.

\textbf{The basic usage of ChatGPT may not produce a desirable performance for the text-to-testcase generation task.} 
When comparing the \texttt{GPT3.5FineTune+Imprv.P} model and the \texttt{GPT3.5 NoFineTune+Basic.P} model, we found that the basic usage of ChatGPT (GPT-3.5-turbo with no fine-tuning and a basic prompt) may not achieve a desirable performance. 

We find that our proposed approach (i.e., the fine-tuned GPT-3.5-turbo with an effective prompting design) is able to generate test cases that are 78.5\% being syntactically correct, 67.09\% being aligned with the requirements while achieving a code coverage of 61.7\%, e, and 18.9\% mutation score which substantially outperforms all other LLMs). 
These results highlight the significant advancement of our approach for the text-to-testcase generation task when compared to other baseline LLMs.

\revise{R1.7, 23.9}{Our approach \texttt{GPT3.5FineTune+Imprv.P} model emerges as the strongest model across all projects and evaluation metrics, particularly excelling in syntax correctness, requirement alignment, code coverage, and mutation scores.
Gemini, while the next best performer, shows commendable results that are 72.35\% syntactically correct, 62.94\% aligned with the requirements while achieving a code coverage of 54.76\% and a mutation score of  15.3\%.
Additionally, GPT-3.5+FineTunetUniTest incurs a cost for generating test cases. Conversely, for projects requiring more cost-effective solutions, Gemini presents a viable alternative that still maintains a comparable level of performance across all key metrics.
The other models CodeGemma and CodeLlama are also consistently strong performers, particularly in \texttt{FineTune+Imprv.P} settings.
}

When fine-tuned the LLMs (GPT-3.5-turbo) with our curated fine-tuned dataset, the performance is improved for all evaluation aspects when compared to the basic usage of ChatGPT. With our improved prompts, the performance is improved for all evaluation aspects when compared to the basic usage of ChatGPT.

When comparing the models between \texttt{GPT3.5 NoFine-Tune + Basic.P} and \texttt{GPT3.5 Fine-Tune + Basic.P}, we observe an improvement of the syntax correctness by 223\%, the requirements alignment by 164\%, and the code coverage by 153\% (see Figure~\ref{ablation}).
In terms of syntax correctness, we found that the performance is improved from 12.90\% for the GPT-3.5 model without fine-tuning \texttt{GPT3.5 NoFine-Tune + Basic.P} to 41.60\% for the GPT-3.5 model with fine-tuning \texttt{GPT3.5 Fine-Tune + Basic.P}.

In terms of requirement alignment, we found that performance improves from 11.10\% for the GPT-3.5 model without fine-tuning \texttt{GPT3.5 NoFine-Tune + Basic.P} to 29.25\% for the GPT-3.5 model with fine-tuning \texttt{GPT3.5 Fine-Tune + Basic.P}.
In terms of code coverage, we found that performance is improved from 10.44\% for the GPT-3.5 model without fine-tuning \texttt{GPT3.5 NoFine-Tune + Basic.P} to 26.23\% for the GPT-3.5 model with fine-tuning \texttt{GPT3.5 Fine-Tune + Basic.P}.
These results confirm that the fine-tuning step with our own curated dataset substantially improves the performance of GPT-3.5, suggesting that model fine-tuning is a crucial step when using a large language model like GPT-3.5, which should be considered in the future.

\begin{custombox}
Our proposed approach (i.e., the fine-tuned GPT-3.5-turbo with an effective prompting design) is able to generate test cases that are 78.5\% being syntactically correct, 67.09\% being aligned with the requirements while achieving a code coverage of 61.7\% and mutation score with 18.9\%, which substantially outperforms all other LLMs. 
\end{custombox}

\revise{R1.3, R2.5}{\subsection*{\textbf{RQ2: \rqtwo}}

\begin{figure}
  \centering
  \includegraphics[width=\linewidth]{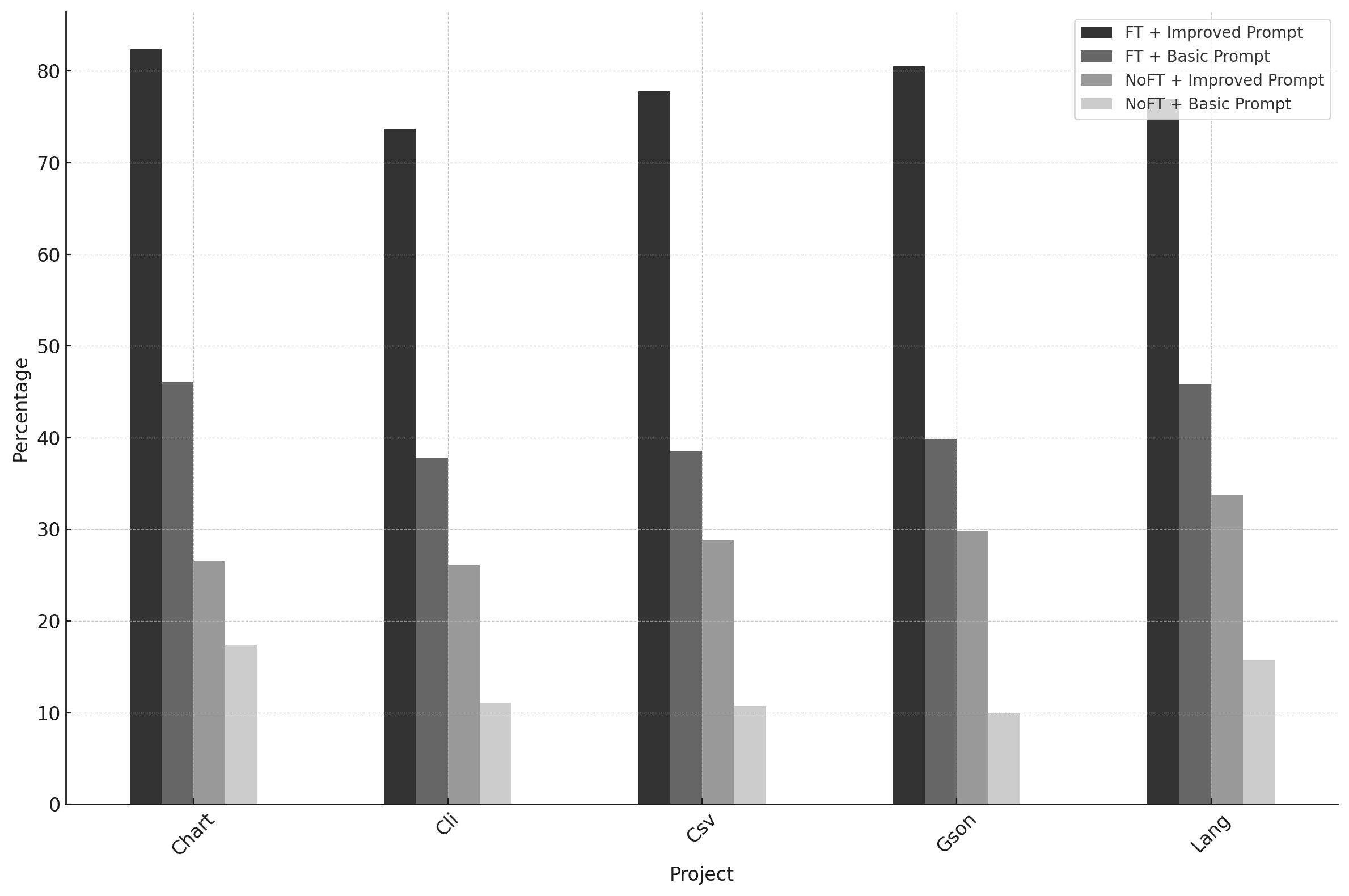}
  \caption{(RQ2)The experimental results of the ablation study with GPT3.5 model}
  \label{ablation}
\end{figure}

When comparing the models between \texttt{GPT3.5 NoFine-Tune + Basic.P} and \texttt{GPT3.5 Fine-Tune + Basic.P}, we observe an improvement of the syntax correctness by 223\%, the requirements alignment by 164\%, and the code coverage by 153\% (see figure~\ref{ablation}).
In terms of syntax correctness, we found that the performance is improved from 12.90\% for the GPT-3.5 model without fine-tuning \texttt{GPT3.5 NoFine-Tune + Basic.P} to 28.90\% for the GPT-3.5 model with fine-tuning \texttt{GPT3.5 Fine-Tune + Basic.P}.
In terms of requirement alignment, we found that the performance is improved from 11.10\% for the GPT-3.5 model without fine-tuning \texttt{GPT3.5 NoFine-Tune + Basic.P} to 20.25\% for the GPT-3.5 model with fine-tuning \texttt{GPT3.5 Fine-Tune + Basic.P}.
In terms of code coverage, we found that the performance is improved from 10.44\% for the GPT-3.5 model without fine-tuning \texttt{GPT3.5 NoFine-Tune + Basic.P} to 16.43\% for the GPT-3.5 model with fine-tuning \texttt{GPT3.5 Fine-Tune + Basic.P}.

When comparing the models between GPT3.5 NoFine-Tune + Basic.P and GPT3.5 NoFine-Tune +Imprv.P, GPT3.5FineTune+Imprv.P and GPT3.5FineTune+Basic.P.
Figure~\ref{ablation} illustrates that optimal results are achieved through the integration of fine-tuning and enhanced prompt engineering with GPT-3.5. It underscores the necessity of incorporating both fine-tuning and effective prompt design as essential strategies when deploying large language models for text-to-testcase generation tasks.

\revise{R2.8}{We employ the Wilcoxon Signed-Rank Test to evaluate the impact of fine-tuning and prompting on the performance metrics syntax correctness, requirement alignment, code coverage and mutation score(SC\_Score, RA\_Score, CC\_Score, MS\_Score). This non-parametric test is used to assess the differences in the median values. We use the test to estimate the relative contribution and the statistical significance (p-value).
The tests compare the metrics under two conditions:

With and without fine-tuning comparison assesses whether the fine-tuning of the model significantly affects the performance metrics.
With and without prompting comparison evaluates the effectiveness of prompting strategies in enhancing model performance.
The results indicate statistically significant differences in all metrics for both conditions, with p-values significantly less than 0.05. This suggests that both fine-tuning and prompting contribute positively to improving model performance across various metrics.
}

\begin{table}[h]
\centering
\caption{Wilcoxon Signed-Rank Test results for model performance.}
\label{tab:wilcoxon_results}
\begin{tabular}{|c|c|c|c|}
\hline
\textbf{Comparison} & \textbf{Metric} & \textbf{Statistic} & \textbf{P-value} \\ \hline
\multirow{4}{*}{Prompting} & SC\_Score & 21.0 & 0.0021 \\ \cline{2-4} 
                             & RA\_Score & 25.0 & 0.0027 \\ \cline{2-4} 
                             & CC\_Score & 33.0 & 0.0133 \\ \cline{2-4} 
                             & MS\_Score & 24.0 & 0.0011 \\ \hline
\multirow{4}{*}{Fine-Tuning }   & SC\_Score & 23.0 & 0.0012 \\ \cline{2-4} 
                             & RA\_Score & 27.0 & 0.0023 \\ \cline{2-4} 
                             & CC\_Score & 40.0 & 0.0136 \\ \cline{2-4} 
                             & MS\_Score & 40.0 & 0.0136 \\ \hline
\end{tabular}
\end{table}

}
\begin{custombox}
When fine-tuning the LLMs (GPT-3.5-turbo) with our curated fine-tuning data set, we observed an improvement of syntax correctness by 223\%, the alignment of the requirements by 164\%,  the coverage of the code by 153\%, and mutation score improved by 273\% compared to the GPT-3.5-turbo model without the fine-tuning component. When considering our improved prompt design, we observed an improvement in syntax correctness by 124\%, requirement alignment by 82\%,  code coverage by 58\%, and mutation score improved by 150\% compared to the GPT-3.5-turbo model without the fine-tuning component.
\end{custombox}

\subsection*{\textbf{RQ3: \rqthree}}

\textbf{Results.} \textbf{The majority of the generated test cases by our approach are syntactically correct for 78.5\%, while being incorrect for 21.5\%. The most common types of errors are assertion errors (11.3\%), value errors (2.4\%), syntax errors (6.9\%), and others (0.9\%).} 

\begin{figure}
  \centering
  \includegraphics[width=\linewidth]{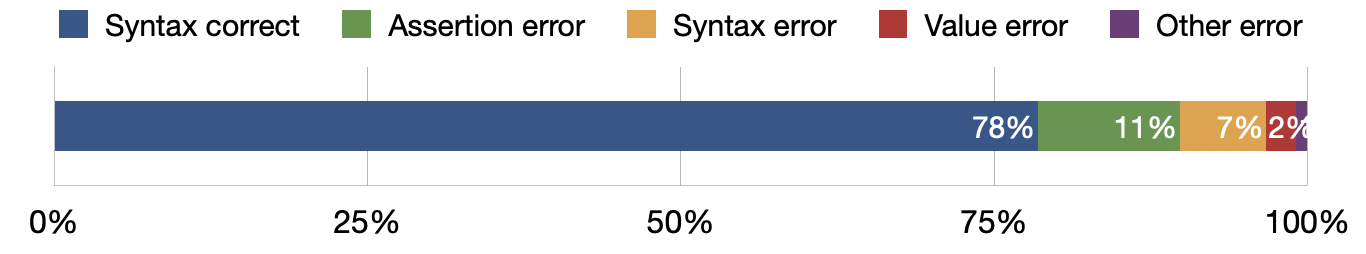}
  \caption{(RQ4)The common types of errors that occur in the generated test cases.}
  \label{error}
\end{figure}

Understanding the incorrectly generated test cases by our approach is crucial as it helps researchers gain a better understanding of the limitations of our approach.
To do so, we utilized Apache Maven~\citep{maven} as our build tool, and the Surefire Plugin~\citep{surefire} to execute the test cases and to analyze the types of errors during the build process. 
Figure~\ref{error} provides a statistical summary of the error types of the generated test cases.
An example of different types of incorrectly generated test cases is also presented in Listing~\ref{FocalClass}.
Below, we discuss the different types of errors.

\begin{itemize}
\item Assertion errors (11.3\%) occur when an assertion statement (a statement that checks if a condition is true) fails because the condition was false. 
The results suggest that a noticeable fraction of the test cases had issues with the logical conditions expected to be true during the tests. 
The assertion error "Escape character should match the default escape character" indicates a failure in a JUnit test case.

\item Syntax errors (6.9\%) refer to errors that do not conform to the language rules.
As a result, the generated test cases cannot be parsed or compiled. For example, the syntax error in the provided code is due to the missing opening curly brace \texttt{'\{'} to begin the method body after the method declaration. 
The code should have an opening brace \texttt{'\{'} after the method declaration to define the scope of the method and include the test logic inside it.

\item A smaller proportion of the test cases (2.4\%) had issues with passing due to inappropriate or unexpected values. For example, in the ``Value error'' indicated by ``junit.framework.ComparisonFailure,'' it means that the test case failed because the actual line separator returned by \texttt{format.getLineSeparator()} did not match the expected line separator (``\textbackslash n'') specified in the test.

\revise{R1.6}{\item The other (0.9\%) category includes exception errors and runtime errors. 
The exception errors occur when the test case throws an unexpected exception or fails to handle an expected exception, such as \texttt{NullPointerException}, \texttt{IndexOutOfBoundsException}, \texttt{IllegalArgumentException}. 
The runtime errors also occur during the execution of the test case, such as \texttt{NullPointerException} or \texttt{IllegalStateException}.}

\end{itemize}

\begin{lstlisting}[float, language = Java , escapeinside={(*@}{@*)}, , caption= Types of Error, label=FocalClass]
// Assertion error -junit.framework.AssertionFailedError: Escape character should match the default escape character
@Test
public void testGetEscape() {
    CSVFormat format = CSVFormat.DEFAULT;
    char expectedEscape = '\\';
    assertEquals("Escape character should match the default escape character", expectedEscape, format.getEscape());}

// Syntax error - java: <identifier> expected, missing '{' in the method
@Test
public void testIsSurroundingSpacesIgnored()
    CSVFormat format = CSVFormat.DEFAULT;
    boolean spacesIgnored = format.isSurroundingSpacesIgnored();
    assertTrue("Surrounding spaces should be ignored by default", spacesIgnored);}

// Value error -junit.framework.ComparisonFailure: Line separator should match the default line separator
@Test
public void testGetLineSeparator() {
    CSVFormat format = CSVFormat.DEFAULT;
    String expectedLineSeparator = "\n";
    assertEquals("Line separator should match the default line separator", expectedLineSeparator, format.getLineSeparator());}
// Other error - Try to access value, which is null, leading to a NullPointerException
@Test
public void testJsonNullConstructor() {
        JsonNull jsonNull = new JsonNull();
        assertNotNull(jsonNull);
        String value = jsonNull.getValue(); 
        assertEquals("null", value);}
\end{lstlisting}

\begin{custombox}
found that a significant portion of the test cases generated by our approach exhibited syntactical correctness, with 78.5\% of the generated test cases are syntactically correct, while 21.5\% of the generated test cases are incorrect.
Among the 21. 5\% of incorrectly generated test cases, we found that the errors are related to assertion errors for 11.3\%, value errors for 2.4\%, syntax errors for 6.9\%, and others for 0.9\%.
\end{custombox}

\section{Discussion}~\label{sec:Discussion}
\subsection{Recommendations to practitioners.}
 
\revise{R2.6}{Consider Alice, a software developer engaged in a project that embraces Test-Driven Development (TDD) practices. In TDD, developers first write a failing test that defines a desired improvement or new function, then produce code to pass the test, and finally refactor the new code to acceptable standards. While this process inherently boosts software quality and aligns development with user requirements, it is also labor-intensive and demands high accuracy in test creation. Unfortunately, most automated test case generation tools available are designed primarily to work with existing code rather than generating tests from scratch based on new requirements. This presents a significant challenge and points to the need for tools capable of generating test cases directly from textual requirements descriptions.
To bridge this gap, practitioners are encouraged to employ Large Language Models (LLMs) specifically fine-tuned for Text-to-Testcase generation tasks. 
The effectiveness of such models is notably enhanced when they are fine-tuned and utilized with well-crafted prompts. The results of RQ2 shows that the basic GPT-3.5 models, when fine-tuned and used with improved prompts, significantly outperform those that are not fine-tuned and employ basic prompts. 
By integrating these advanced LLMs into their workflow, developers can enhance their efficiency in test case generation, thereby adhering more effectively to TDD practices.}

\subsection{Usability and practicality.}

\begin{figure}
  \centering
  \includegraphics[width=\linewidth]{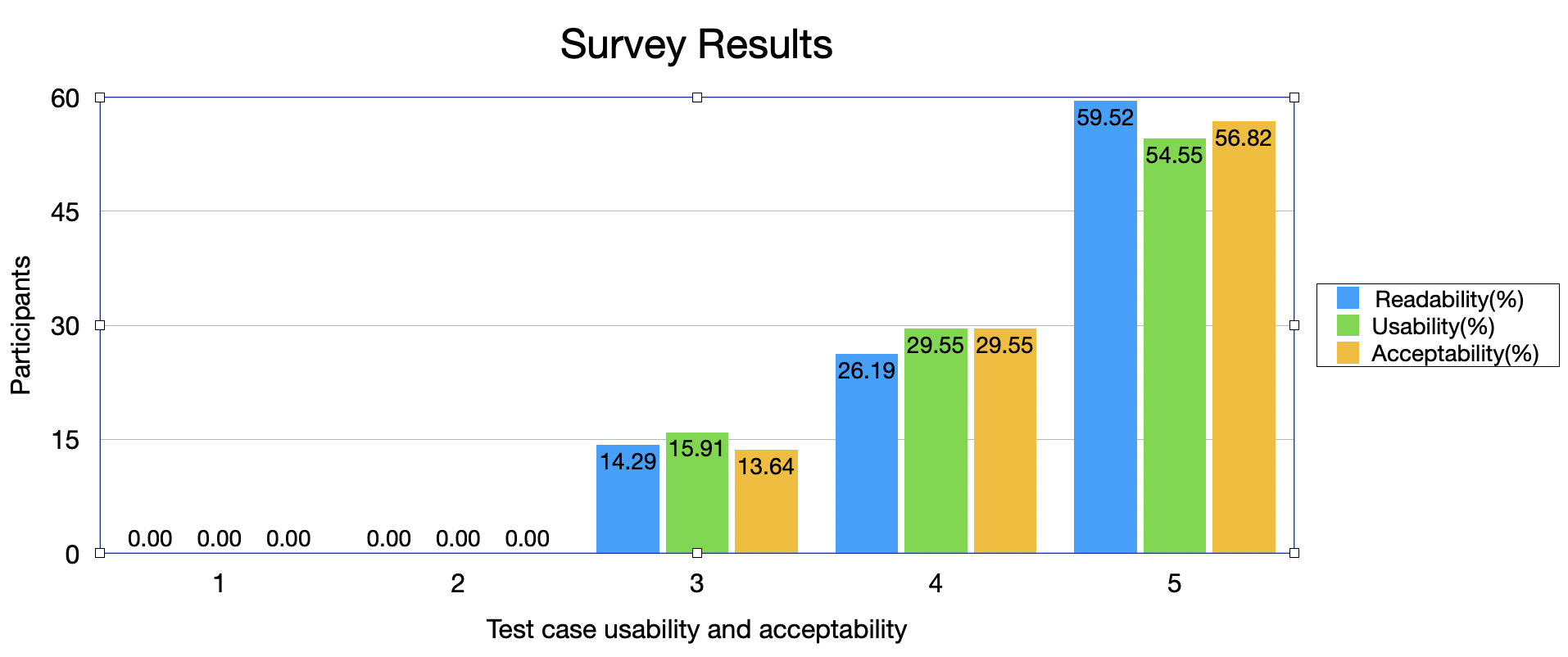}
  \caption{Survey Results of usability and acceptability of the test cases generated.}
  \label{SurveyResults}
\end{figure}

\revise{R3.5}{To understand the usability and acceptability of the test cases generated by our approach to text-to-testcase generation, we conducted a user study. We aim to unveil the usability and acceptability of the test cases generated by our approach. 
We conduct our user study according to the following steps: (1) design and develop a survey, (2) select participants, and (3) survey questions.
We explain the details of each step below.

(i) Design and Development of the Survey: Our survey utilises a cross-sectional design, capturing responses from participants at a singular point in time. 
Comprising four closed-ended and one open-ended question, the survey is structured to be completed anonymously in approximately 20 minutes. 
The closed-ended questions use a Likert scale format the scale ranges from 1 (strongly disagree) to 5 (strongly ggree), where each number corresponds to a specific level of agreement or satisfaction, and the survey itself is divided into two sections: initial demographic questions and questions regarding the usability and acceptability of test cases generated by our approach.
We employed Qualtrics as the platform for our online survey administration, participants encountered a detailed introductory statement upon entering the survey, which outlines the study’s goals, the basis for participant selection, potential benefits and risks, and our dedication to maintaining confidentiality. 
Importantly, our survey underwent a rigorous evaluation process and received ethical approval from the Monash University Human Research Ethics Committee (MUHREC ID: 45001).

During the survey, participants were presented with test cases generated by the fine-tuned GPT3.5 our approach. 
For the five open-source projects we randomly selected two test cases from each project, resulting in a total of 10 test cases. 
Participants were asked to evaluate the usability and acceptability of each test case by answering four questions: the first focused on determining which test case was easier to read; the second aimed to identify relevance and correctness of the test case; the third asked participants to express their willingness to accept and adopt the generated test case for the given method description in one of their own projects, considering its relevance and quality; the fourth focus on the any other feedback or comments.
(ii) Participant Selection: The target population for our study is Java software practitioners. We reached out to potential participants through LinkedIn and Facebook. Participants were invited to complete the online survey, which assessed the readability,usability and acceptability of test cases. Finally, we obtained a total of 61 responses in two weeks.

(iii) Survey Questions:
The survey begins with a question about the participant's role within their software development team ("(Q1). What is your role in the software development team?"). The next question concerns the years of experience the participants have in Java programming ("(Q2). How many years of experience do you have in Java programming?"), aimed at achieving a diverse range of responses from software practitioners with varied levels of professional experience.

The survey included precisely four questions presented to the participants: "(Q1) How would you rate the readability of the generated test case?"; followed by (Q2) "How would you rate the usability of the generated test cases?"; additionally (Q3) "How would you to rank the acceptability of the generated test cases?"; finally any other feedback or comments.

To ensure the completeness of survey responses, especially for open-ended questions, we conducted a detailed manual review. Out of a total of 61 responses, we identified 8 invalid responses—due to unanswered open-ended questions or incomprehensible replies—and excluded them from further analysis. Consequently, we included the remaining 53 valid responses for analysis. The closed-ended responses were quantitatively analyzed. 

The figure~\ref{SurveyResults} presents the results of a survey using a 5-point Likert scale to assess the readability, usability, and acceptability of the generated test cases. 
The Likert scale ranges from 1 (Strongly Disagree or Very Poor) to 5 (Strongly Agree or Excellent). 
No respondents were rated Very Poor or Poor (ratings 1 and 2), indicating that there was no negative feedback. 
A minority of the respondents were neutral (rating 3), with percentages ranging from 14\% to 16\%. 
A larger proportion agreed that the test cases were good (rating 4), with scores ranging from about 26\% to 30\%. 
The majority felt strongly positive, rating the test cases as Excellent (rating 5), with significant majority above 50\%. 
This distribution suggests a general positive reception for the usability and acceptability of text-to-testcase generation.

}

\section{Threats to Validity}~\label{sec:Threats}
Below, we discuss the most pertinent threats to the validity of our study.

\sectopic{Internal Validity.} 
The main internal validity concerns in our study are two-fold: (i) hyper-parameter optimization of various algorithms and (ii) the potential training data-leakage issue of LLMs. 

In our fine-tuning step, we employed a variety of hyperparameter configurations, including adjustments in the number of epochs, batch size, and learning rate. The hyperparameter optimization can significantly improve 
evaluation outcomes~\citep{bischl2023hyperparameter}. 

However, identifying the optimal settings is resource-intensive due to the expansive search space of hyperparameters. We note that our primary objective in this study is not to pinpoint the most effective hyperparameters but to investigate the potential of LLMs in TDD. Thus, while the results achieved with LLMs in our study are promising, they only represent a starting point and have the potential for further improvement through extensive hyperparameter tuning. Nonetheless, we have included detailed information about the hyperparameter settings used in our study within our replication package. The second major concern related to LLMs is the training data leakage~\citep{xia2023automated}. There is a possibility that GPT-3.5-turbo or other LLMs may have been exposed to the same open-source projects that we used in our study. However, the pre-trained models did not yield satisfactory results, compared to the fine-tuned model and custom prompt results. This shows that the impact of training data leakage, if any, was minimal in our study settings.

\sectopic{Construct Validity.} 
Our study's main construct validity concern is related to the prompt tuning and robustness~\citep{lester2021power}. Given the large scope of subjectivity in prompting and the lack of empirical guidelines on prompt tuning and robustness in SE, one can always further tune and build a more robust prompt to achieve better evaluation results. However, we experimented with several prompt combinations in our journey, from the basic to the custom prompt, in an attempt to mitigate the threats related to prompt tuning.
\revise{R1.5}{The second major concern is related to the potential data leakage (i.e., when instances in the training dataset appear in the evaluation set) of our model. To mitigate this threat, we focus on identifying instances in the training dataset (i.e., method2test) that appear in the evaluation set (i.e., open-source projects). For the method2test dataset, we discovered and removed 63 instances for the JFreeChart project. This will ensure that there is no data leakage in our experiments.}

\sectopic{External Validity} 
The external validity concerns the generalization of our findings. We experimented with five open source Java projects from Github, selected with diversity in mind to improve our external validity. 
The five projects represent different domains of inputs (string, int, etc.) and sizes and complexity of classes. 
Further experiments with other open-source projects would help with the generalizability of the results.

\section{Conclusion}~\label{sec:Conclusion}
In this paper, we introduce a Text-to-Testcase generation approach based on a fine-tuned large language model with an effective prompting design.
Our results show that the prompting-based fine-tuning GPT3.5 outperforms other baselines. 
Furthermore, our ablation study demonstrates the substantial performance improvement of the fine-tuning and prompting components of the GPT3.5 model, highlighting the importance of the fine-tuning and prompting components for the text-to-testcase generation task.
Based on our results, we draw the following recommendations: (1) the basic usage of ChatGPT may not produce a desirable performance for the text-to-testcase generation task; and (2) both fine-tuning and effective prompting design should always be considered when using a large language model.

\bibliographystyle{elsarticle-num}

\bibliography{sample-base}

\begin{thebibliography}{10}
\expandafter\ifx\csname url\endcsname\relax
  \def\url#1{\texttt{#1}}\fi
\expandafter\ifx\csname urlprefix\endcsname\relax\def\urlprefix{URL }\fi
\expandafter\ifx\csname href\endcsname\relax
  \def\href#1#2{#2} \def\path#1{#1}\fi

\bibitem{beck2003test}
K.~Beck, Test-driven development: by example, Addison-Wesley Professional, 2003.

\bibitem{madeyski2010test}
L.~Madeyski, G.~de~sistemas~de informaci{\'o}n, Test-driven development: An empirical evaluation of agile practice, Springer, 2010.

\bibitem{barraood2021comparison}
S.~O. Barraood, H.~Mohd, F.~Baharom, A comparison study of software testing activities in agile methods, in: Knowledge Management International Conference (KMICe), Vol. 2021, University Utara Malaysia Kedah, Malaysia, 2021.

\bibitem{wasmus2007evaluation}
H.~Wasmus, H.-G. Gross, Evaluation of test-driven development-an industrial case study, in: International Conference on Evaluation of Novel Approaches to Software Engineering, Vol.~2, SCITEPRESS, 2007, pp. 103--110.

\bibitem{maximilien2003assessing}
E.~M. Maximilien, L.~Williams, Assessing test-driven development at ibm, in: 25th International Conference on Software Engineering, 2003. Proceedings., IEEE, 2003, pp. 564--569.

\bibitem{beck2022test}
K.~Beck, Test driven development: By example, Addison-Wesley Professional, 2022.

\bibitem{pacheco2007randoop}
C.~Pacheco, M.~D. Ernst, Randoop: feedback-directed random testing for java, in: Companion to the 22nd ACM SIGPLAN conference on Object-oriented programming systems and applications companion, 2007, pp. 815--816.

\bibitem{fraser2011evosuite}
G.~Fraser, A.~Arcuri, Evosuite: automatic test suite generation for object-oriented software, in: Proceedings of the 19th ACM SIGSOFT symposium and the 13th European conference on Foundations of software engineering, ACM, 2011, pp. 416--419.

\bibitem{tufano2020unit}
M.~Tufano, D.~Drain, A.~Svyatkovskiy, S.~K. Deng, N.~Sundaresan, Unit test case generation with transformers and focal context, arXiv preprint arXiv:2009.05617 (2020).

\bibitem{alagarsamy2023a3test}
S.~Alagarsamy, C.~Tantithamthavorn, A.~Aleti, A3test: Assertion-augmented automated test case generation, arXiv preprint arXiv:2302.10352 0~(0) (2023) 0.

\bibitem{chen2021evaluating}
M.~Chen, J.~Tworek, H.~Jun, Q.~Yuan, H.~P. D.~O. Pinto, J.~Kaplan, H.~Edwards, Y.~Burda, N.~Joseph, G.~Brockman, et~al., Evaluating large language models trained on code, arXiv preprint arXiv:2107.03374 (2021).

\bibitem{roziere2023code}
B.~Roziere, J.~Gehring, F.~Gloeckle, S.~Sootla, I.~Gat, X.~E. Tan, Y.~Adi, J.~Liu, T.~Remez, J.~Rapin, et~al., Code llama: Open foundation models for code, arXiv preprint arXiv:2308.12950 (2023).

\bibitem{svyatkovskiy2019pythia}
A.~Svyatkovskiy, Y.~Zhao, S.~Fu, N.~Sundaresan, Pythia: Ai-assisted code completion system, in: Proceedings of the 25th ACM SIGKDD International Conference on Knowledge Discovery \& Data Mining, 2019, pp. 2727--2735.

\bibitem{tufano2019empirical}
M.~Tufano, C.~Watson, G.~Bavota, M.~D. Penta, M.~White, D.~Poshyvanyk, An empirical study on learning bug-fixing patches in the wild via neural machine translation, ACM Transactions on Software Engineering and Methodology (TOSEM) 28~(4) (2019) 1--29.

\bibitem{chen2019sequencer}
Z.~Chen, S.~Kommrusch, M.~Tufano, L.-N. Pouchet, D.~Poshyvanyk, M.~Monperrus, Sequencer: Sequence-to-sequence learning for end-to-end program repair, IEEE Transactions on Software Engineering 47~(9) (2019) 1943--1959.

\bibitem{hu2020deep}
X.~Hu, G.~Li, X.~Xia, D.~Lo, Z.~Jin, Deep code comment generation with hybrid lexical and syntactical information, Empirical Software Engineering 25~(3) (2020) 2179--2217.

\bibitem{watson2020learning}
C.~Watson, M.~Tufano, K.~Moran, G.~Bavota, D.~Poshyvanyk, On learning meaningful assert statements for unit test cases, in: Proceedings of the ACM/IEEE 42nd International Conference on Software Engineering, 2020, pp. 1398--1409.

\bibitem{openai2023gpt}
R.~OpenAI, Gpt-4 technical report. arxiv 2303.08774, View in Article (2023).

\bibitem{luccioni2022estimating}
A.~S. Luccioni, S.~Viguier, A.-L. Ligozat, Estimating the carbon footprint of bloom, a 176b parameter language model, arXiv preprint arXiv:2211.02001 (2022).

\bibitem{wang2021codet5}
Y.~Wang, W.~Wang, S.~Joty, S.~C. Hoi, Codet5: Identifier-aware unified pre-trained encoder-decoder models for code understanding and generation, arXiv preprint arXiv:2109.00859 (2021).

\bibitem{fried2022incoder}
D.~Fried, K.~Agrawal, E.~Ho, et~al., Incoder: A generative model for code infilling and synthesis, in: NeurIPS, 2022.

\bibitem{li2023starcoder}
R.~Li, L.~B. Allal, Y.~Zi, N.~Muennighoff, D.~Kocetkov, C.~Mou, M.~Marone, C.~Akiki, J.~Li, J.~Chim, et~al., Starcoder: may the source be with you!, arXiv preprint arXiv:2305.06161 (2023).

\bibitem{team2024codegemma}
C.~Team, Codegemma: Open code models based on gemma, arXiv preprint arXiv:2406.11409 (2024).

\bibitem{gemini2024}
G.~DeepMind, Gemini: Scaling vision-llms with efficient training and inference (2024).
\newblock \href {http://arxiv.org/abs/2401.01101} {\path{arXiv:2401.01101}}.

\bibitem{Apache}
A.~C. Lang (2022).
\newblock \href{https://commons.apache.org/proper/commons-lang/}{[link]}.
\newline\urlprefix\url{https://commons.apache.org/proper/commons-lang/}

\bibitem{JFreeChart}
JFreeChart (2022).
\newblock \href{https://jfree.org/jfreechart/}{[link]}.
\newline\urlprefix\url{https://jfree.org/jfreechart/}

\bibitem{ApacheCommonsCLI}
A.~C. CLI (2022).
\newblock \href{https://commons.apache.org/proper/commons-cli/}{[link]}.
\newline\urlprefix\url{https://commons.apache.org/proper/commons-cli/}

\bibitem{ApacheCommonsCSv}
C.~CSV (2022).
\newblock \href{https://commons.apache.org/proper/commons-csv/}{[link]}.
\newline\urlprefix\url{https://commons.apache.org/proper/commons-csv/}

\bibitem{googleGson}
G.~Gson (2019).
\newblock \href{https://github.com/google/gson/}{[link]}.
\newline\urlprefix\url{https://github.com/google/gson/}

\bibitem{astels2003test}
D.~Astels, Test driven development: A practical guide, Prentice Hall Professional Technical Reference, 2003.

\bibitem{mueller2002experiment}
M.~Mueller, A.~Zeller, An experiment in test-driven development: Lessons learned, in: Proceedings of the 2002 ACM SIGSOFT International Symposium on Foundations of Software Engineering, ACM, 2002, pp. 238--246.

\bibitem{janzen2008does}
D.~Janzen, Does test-driven development really improve software quality?, IEEE Computer 41~(5) (2008) 79--84.

\bibitem{erdogmus2005effectiveness}
H.~Erdogmus, S.~Arisan, The effectiveness of test-driven development technique in reducing defect density, Journal of Research and Practice in Information Technology 38~(2) (2005) 185--191.

\bibitem{Jurafsky:20}
D.~Jurafsky, J.~H. Martin, Speech and Language Processing, 3rd Edition, 2020.

\bibitem{wei2022emergent}
J.~Wei, Y.~Tay, R.~Bommasani, C.~Raffel, B.~Zoph, S.~Borgeaud, D.~Yogatama, M.~Bosma, D.~Zhou, D.~Metzler, et~al., Emergent abilities of large language models, arXiv preprint arXiv:2206.07682 (2022).

\bibitem{nguyen2023generative}
A.~Nguyen-Duc, B.~Cabrero-Daniel, A.~Przybylek, C.~Arora, D.~Khanna, T.~Herda, U.~Rafiq, J.~Melegati, E.~Guerra, K.-K. Kemell, et~al., Generative artificial intelligence for software engineering--a research agenda, arXiv preprint arXiv:2310.18648 (2023).

\bibitem{hou2023large}
X.~Hou, Y.~Zhao, Y.~Liu, Z.~Yang, K.~Wang, L.~Li, X.~Luo, D.~Lo, J.~Grundy, H.~Wang, Large language models for software engineering: A systematic literature review, arXiv preprint arXiv:2308.10620 (2023).

\bibitem{arora2023advancing}
C.~Arora, J.~Grundy, M.~Abdelrazek, Advancing requirements engineering through generative ai: Assessing the role of llms, arXiv preprint arXiv:2310.13976 (2023).

\bibitem{fan2023large}
A.~Fan, B.~Gokkaya, M.~Harman, M.~Lyubarskiy, S.~Sengupta, S.~Yoo, J.~M. Zhang, Large language models for software engineering: Survey and open problems, arXiv preprint arXiv:2310.03533 (2023).

\bibitem{fu2023chatgpt}
M.~Fu, C.~Tantithamthavorn, V.~Nguyen, T.~Le, Chatgpt for vulnerability detection, classification, and repair: How far are we?, in: 30th Asia-Pacific Software Engineering Conference (APSEC 2023), 2023.

\bibitem{yuan2023no}
Z.~Yuan, Y.~Lou, M.~Liu, S.~Ding, K.~Wang, Y.~Chen, X.~Peng, No more manual tests? evaluating and improving chatgpt for unit test generation, arXiv preprint arXiv:2305.04207 (2023).

\bibitem{brown2020language}
T.~Brown, B.~Mann, N.~Ryder, M.~Subbiah, J.~D. Kaplan, P.~Dhariwal, A.~Neelakantan, P.~Shyam, G.~Sastry, A.~Askell, et~al., Language models are few-shot learners, Advances in neural information processing systems 33 (2020) 1877--1901.

\bibitem{xie2023chatunitest}
Z.~Xie, Y.~Chen, C.~Zhi, S.~Deng, J.~Yin, Chatunitest: a chatgpt-based automated unit test generation tool, arXiv preprint arXiv:2305.04764 (2023).

\bibitem{bissi2016effects}
W.~Bissi, A.~G. S.~S. Neto, M.~C. F.~P. Emer, The effects of test driven development on internal quality, external quality and productivity: A systematic review, Information and Software Technology 74 (2016) 45--54.

\bibitem{wang2023software}
J.~Wang, Y.~Huang, C.~Chen, Z.~Liu, S.~Wang, Q.~Wang, Software testing with large language model: Survey, landscape, and vision, arXiv preprint arXiv:2307.07221 (2023).

\bibitem{schafer2023empirical}
M.~Sch{\"a}fer, S.~Nadi, A.~Eghbali, F.~Tip, An empirical evaluation of using large language models for automated unit test generation, arXiv preprint arXiv:2302.06527 (2023).

\bibitem{siddiq2023exploring}
M.~L. Siddiq, J.~Santos, R.~H. Tanvir, N.~Ulfat, F.~A. Rifat, V.~C. Lopes, Exploring the effectiveness of large language models in generating unit tests, arXiv preprint arXiv:2305.00418 (2023).

\bibitem{dakhel2024effective}
A.~M. Dakhel, A.~Nikanjam, V.~Majdinasab, F.~Khomh, M.~C. Desmarais, Effective test generation using pre-trained large language models and mutation testing, Information and Software Technology 171 (2024) 107468.

\bibitem{yang2024empirical}
L.~Yang, C.~Yang, S.~Gao, W.~Wang, B.~Wang, Q.~Zhu, X.~Chu, J.~Zhou, G.~Liang, Q.~Wang, et~al., An empirical study of unit test generation with large language models, arXiv preprint arXiv:2406.18181 (2024).

\bibitem{hossain2024togll}
S.~B. Hossain, M.~Dwyer, Togll: Correct and strong test oracle generation with llms, arXiv preprint arXiv:2405.03786 (2024).

\bibitem{andrews2005mutation}
J.~H. Andrews, L.~C. Briand, Y.~Labiche, Is mutation an appropriate tool for testing experiments?, in: Proceedings of the 27th international conference on Software engineering, 2005, pp. 402--411.

\bibitem{cai2005effect}
X.~Cai, M.~R. Lyu, The effect of code coverage on fault detection under different testing profiles, in: Proceedings of the 1st International Workshop on Advances in Model-based Testing, 2005, pp. 1--7.

\bibitem{gopinath2014code}
R.~Gopinath, C.~Jensen, A.~Groce, Code coverage for suite evaluation by developers, in: Proceedings of the 36th international conference on software engineering, 2014, pp. 72--82.

\bibitem{hemmati2015effective}
H.~Hemmati, How effective are code coverage criteria?, in: 2015 IEEE International Conference on Software Quality, Reliability and Security, IEEE, 2015, pp. 151--156.

\bibitem{Jacoco}
J.~C.~C. Library (2019).
\newblock \href{https://www.eclemma.org/jacoco/trunk/index.html}{[link]}.
\newline\urlprefix\url{https://www.eclemma.org/jacoco/trunk/index.html}

\bibitem{coles2016pit}
H.~Coles, T.~Laurent, C.~Henard, M.~Papadakis, A.~Ventresque, Pit: a practical mutation testing tool for java, in: Proceedings of the 25th international symposium on software testing and analysis, 2016, pp. 449--452.

\bibitem{tufano2022methods2test}
M.~Tufano, S.~K. Deng, N.~Sundaresan, A.~Svyatkovskiy, Methods2test: A dataset of focal methods mapped to test cases, arXiv preprint arXiv:2203.12776 (2022).

\bibitem{tree-sitter}
Tree-sitter (2022).
\newblock \href{http://tree-sitter.github.io/tree-sitter}{[link]}.
\newline\urlprefix\url{http://tree-sitter.github.io/tree-sitter}

\bibitem{maven}
maven, \href{https://maven.apache.org/}{Apache maven} (2023).
\newline\urlprefix\url{https://maven.apache.org/}

\bibitem{surefire}
surefire, \href{https://maven.apache.org/surefire-maven-plugin/}{Surefire plugin} (2023).
\newline\urlprefix\url{https://maven.apache.org/surefire-maven-plugin/}

\bibitem{bischl2023hyperparameter}
B.~Bischl, M.~Binder, M.~Lang, T.~Pielok, J.~Richter, S.~Coors, J.~Thomas, T.~Ullmann, M.~Becker, A.-L. Boulesteix, et~al., Hyperparameter optimization: Foundations, algorithms, best practices, and open challenges, Wiley Interdisciplinary Reviews: Data Mining and Knowledge Discovery 13~(2) (2023) e1484.

\bibitem{xia2023automated}
C.~S. Xia, Y.~Wei, L.~Zhang, Automated program repair in the era of large pre-trained language models, in: Proceedings of the 45th International Conference on Software Engineering (ICSE 2023). Association for Computing Machinery, 2023.

\bibitem{lester2021power}
B.~Lester, R.~Al-Rfou, N.~Constant, The power of scale for parameter-efficient prompt tuning, arXiv preprint arXiv:2104.08691 (2021).

\end{thebibliography}

 \end{document}